\documentclass[twocolumn]{aa}
\usepackage{txfonts}
\usepackage{graphicx,isolatin1}
\begin{document}
\title{ISO observations of the Galactic center Interstellar Medium: \thanks{Based on observations with ISO, an ESA project with instruments funded by ESA Member States (especially the PI countries: France, Germany, the Netherlands and the United Kingdom) and with the participation of ISAS and NASA.}  }

\subtitle{ionized gas}

\author{N. J. Rodriguez-Fernandez \inst{1} \fnmsep\thanks{Marie Curie Fellow}
  \and J. Martin-Pintado \inst{2}  }

\offprints{N. J. Rodríguez-Fernández}
 
\institute{LERMA (UMR 8112), Observatoire de Paris, 61 Av de l'Observatoire,
   F-75014 Paris, France\\
              \email{nemesio.rodriguez@obspm.fr}
         \and
         DAMIR, Instituto de Estructura de la Materia, Consejo Superior de Investigaciones Científicas (CSIC), Serrano 121, E-28006 Madrid, Spain
             }
\date{Received ; accepted}

\abstract{
We present fine structure and recombination lines observations of the ionized gas toward a sample of 18 sources located within 300 pc of the center of the Galaxy (hereafter Galactic center, GC). The sources were selected as molecular clouds located far from thermal continuum sources. The fine structure lines from [N{\sc ii}] and [S{\sc iii}] have been detected in 16 sources. In 10 sources we have even detected the [O{\sc iii}] 88 $\mu$m line. Several techniques have been used to determine lower and upper limits to the extinction toward each source to correct the observed line fluxes. The derived electron densities of the ionized gas vary from  $\sim 100$ to $\leq 30$ cm$^{-3}$. For some sources we were able to  derive N, S and Ne abundances. We  found that they are similar to those measured in the H{\sc ii} regions in the 5-kpc ring and in the nuclei of starburst galaxies.  The fine structure  lines ratios measured for all the sources can be explained by photo-ionization with an effective temperature of the ionizing radiation of 32000-37000 K and an ionization parameter, $U$, of $-1>\log U > -3$.  The highest excitation  is found in the Radio Arc region but it does not decrease smoothly with distance. There must be more ionizing sources distributed over the Galactic center than the known clusters of massive stars. Most of the clouds are located far (up to 45 pc for M-0.96+0.13) from the prominent continuum complexes (Sgr C, B ...). However, it is possible that the clouds are ionized by escaped photons from those complexes. The comparison of the effective temperatures of the ionizing radiation to the measured Lyman continuum photons emission rate imply that the clouds are indeed ionized by distant sources. The excitation ratios, effective temperature and ionization parameter measured in the GC are similar to those found in some low excitation starburst galaxies. The [Ne{\sc iii}]/[Ne{\sc ii}] line ratios  measured in the GC sources are consistent with the results of the Thornely et al. (2000) model for a short burst of massive star formation less than 8 Myr ago. We have also found that the [Ne{\sc ii}] 13 $\mu$m ~ to far-infrared continuum ratio measured for the GC sources is similar to that of external galaxies, supporting the idea by Sturm et al. (2002) that the far-infrared continuum in Active Galaxies is dominated by dust heated by stellar  radiation rather than by the AGN.

}


\maketitle
%


\def\kms{~km~s$^{-1}$}
\def\cmmt{~cm$^{-3}$}
\def\cmmd{~cm$^{-2}$}
\def\smu{s$^{-1}$}
\def\mum{$\mu$m}
\newcommand{\gsim}{\raisebox{-.4ex}{$\stackrel{>}{\scriptstyle \sim}$}}
\newcommand{\lsim}{\raisebox{-.4ex}{$\stackrel{<}{\scriptstyle \sim}$}}
\def\ls{\lsim}
\def\gs{\gsim}
\def\le{$\leq$}
\def\ge{$\geq$}

\def\HII{H{\sc ii}}

\def\HeII{\mbox{He {\sc ii}}}
\def\NeII{\mbox{Ne{\sc ii}}}
\def\NeIII{\mbox{Ne{\sc iii}}}
\def\NII{\mbox{N{\sc ii}}}
\def\NIII{\mbox{N{\sc iii}}}
\def\OIII{\mbox{O{\sc iii}}}
\def\OI{O{\sc i}}
\def\CII{C{\sc ii}}
\def\SiI{\mbox{Si {\sc i}}}
\def\SiII{\mbox{Si {\sc ii}}}
\def\SIII{\mbox{S {\sc iii}}}
\def\SIV{\mbox{S {\sc iv}}}
\def\ArII{\mbox{Ar {\sc ii}}}
\def\le{$\leq$}


\section{Introduction}

The non-thermal radio emission as well as $\gamma$ and X-rays observations of the Galactic center (hereafter we refer to the 500 central pc of the Galaxy as Galactic center, GC) show the presence of a very hot plasma ($10^7-10^8$  K,  Koyama et al. 1996; Markevitch et al. 1993) and a recent episode of nucleosynthesis  ($\sim10^6$  yr ago;  Diehl et al. 1993)  that suggest an event of violent star formation in the recent past of the GC. On the other hand, the cold medium in the GC is known mainly due to  radio observations of the molecular gas. Up to 10$^8$ M$_\odot$ of molecular gas (10 $\%$ of the neutral gas of the Galaxy) is found in the GC.  This gas is dense (10$^4$ cm$^{-3}$), turbulent (linewidths $\sim$20 \kms) and exhibits relatively high kinetic temperatures from $\sim 15$ to $\sim$150 K (see for instance, Rodríguez-Fernández et al. 2001b).

In between the hot and cold phases, the GC interstellar medium also presents a warm ionized medium that has been studied mainly by radio continuum and hydrogen radio recombination lines observations. For example, in the 5 GHz continuum map by Altenhoff et al. (1978) the ionized gas emission shows an extended component of $500 \times 300$ pc$^2$ and a number of discrete \HII~ regions, most of them associated with the Sgr A, B,... complexes. At this frequency,  Mezger and Pauls (1979) considered that 50$\%$ of the flux is thermal radio continuum to which discrete \HII~ regions contribute with  $\sim 40 \%$ and extended low density (ELD)  ionized gas with $\sim 60 \%$.

The ionized gas  has also been studied by radio recombination lines. Many surveys with low angular resolution (several arc minutes) have been published (see for instance Lockman et al. 1973, Matthew et al. 1973, Pauls and Mezger 1975). Their main results are that the ionized gas rotates in the same direction as the neutral gas and that the ELD ionized gas presents a high degree of turbulence as shown by the large line widths ($\sim 50-100$ \kms). In the last twenty years many observations have been done toward the most prominent  \HII~ regions (Sgr A, B...) and ionized nebulae (The Sickle,...). These are mainly centimeter interferometer or millimeter single-dish observations of hydrogen recombination lines of small regions. Some fine structure lines observations have also been published in the eighties and nineties. Fine structure line ratios are almost independent of the  electron temperature and can be used to derive electron densities and the properties of the ionizing radiation. Using these techniques Lacy et al. (1980) and Genzel et al. (1984) observed broad lines ($\sim 100$ \kms) in the central parsecs of the Galaxy and derived an electron density of $\sim 10^4$ \cmmt ~ and an effective temperature of the ionizing radiation of $\lsim 35000$ K. On the other hand, Simpson et al. (1997) found electron densities of $\sim 200$ \cmmt~ and an effective temperature of the ionizing radiation of 36000 K in the Sickle area.

Recent ISO observations  of fine structure lines have allowed  to study the ELD ionized gas at spatial scales of $\sim 50$ pc in the Radio Arc (Rodriguez-Fernandez et al. 2001a) and in the Sgr B regions (Goicoechea et al. 2004), probing a moderate density (10-100 \cmmt) gas component ionized by diluted and relatively hard star radiation ($\sim$35000 K). In the case of the Radio Arc, the origin of the radiation is the Quintuplet and the Arches clusters while in the case of Sgr B2 the exact location and the characteristics of the ionizing sources are not known (although they must  be located in the vicinity of Sgr B2M).

\begin{figure*}[htb]
\includegraphics[ angle=90,width=18cm]{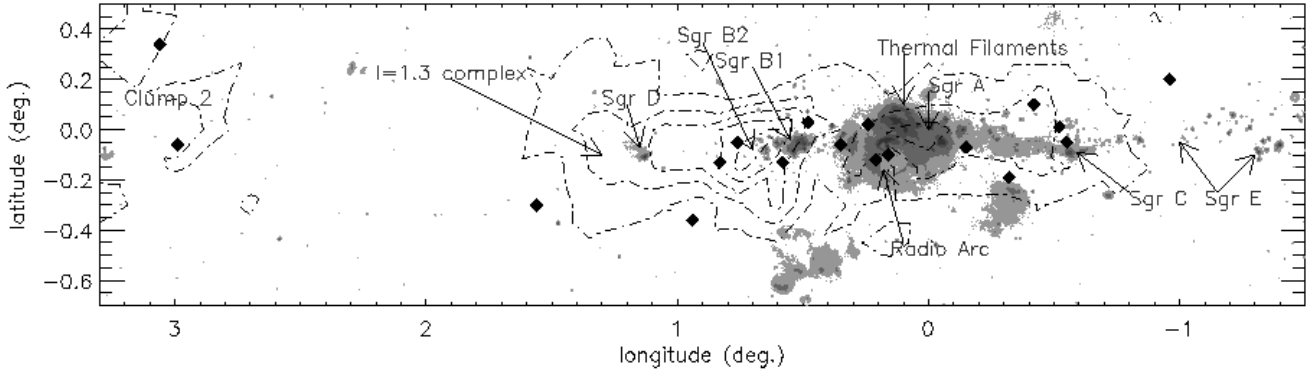}
\caption{Dot-dashed contours represent the C$^{18}$O(1-0) integrated intensity map by Dahmen et al. (1997)  (angular resolution of 8.8$^{'}$.) The 20 \mum~ image by the Midcourse Space Experiment (MSX) is shown in gray levels  (resolution of 9$ ^{''}$, see Egan et al. 1998). The black diamonds  indicate the positions of the sources studied in this paper (a short version of their names is given in black characters). The main molecular and radiocontiuum sources are indicated by arrows and their names are written in gray characters}
\label{fig_msx}
\end{figure*}

 In addition to the detailed studies of outstanding regions of the GC, we have also undertaken a study of the interstellar medium in the 500 central pc of the Galaxy. The goal of the project is to study the thermal balance and the origin of the high temperatures (200 K) of the molecular gas in sources that do not seem to be influenced by stellar radiation (see for instance Hüttemeister et al. 1993). From the CS and SiO large scale maps (Bally et al. 1987, Martin-Pintado et al. 1997, 2000), we have selected a sample of 18 molecular peaks located far from  thermal radiocontinuum or far-infrared sources.  Table \ref{tab_coor} shows the J2000 coordinates of the sources and Fig. \ref{fig_msx} shows the location of the clouds   on  the   20 \mum~ image (gray levels) of the GC taken by the MSX satellite and the C$^{18}$O map (contour lines) by Dahmen et al (1997). The main molecular and continuum emission complexes (Clump 2, Sgr D,..., Sgr E) are also shown in Fig. \ref{fig_msx}. Note that the observed  clouds avoid the continuum peaks. Some of the 18 selected sources were already observed in NH$_3$ lines by  Hüttemeister et al. (1993) while ISO observations of H$_2$ pure-rotational lines toward the 18 sources have been presented by Rodríguez-Fernández et al. (2000, 2001b), who have estimated, for the first time, the total amount of warm molecular gas in the GC clouds. The heating mechanisms of the neutral gas have been studied by Rodríguez-Fernández et al. (2004), who presented fine structure line observations of the neutral gas and the dust continuum spectra of these sources.

In this paper we present infrared fine structure line as well as infrared and radio hydrogen recombination line observations of the ionized gas. . The paper is organized as follows. In Section 2 we present the observations, the data reduction and a comparison between different observations. In Section 3 we discuss the extinction correction. In Sections 4 to 7 we  present the results. The discussion and the conclusions are presented in Sections 8 and 9.

\section{Observations and data reduction}

\begin{table*}[htb]
\caption{J2000 coordinates  of the sources and Target Dedicated Time (TDT) numbers of the ISO observations.}
\label{tab_coor}
\label{tab_isoobs}
\begin{tabular}{llllll}
\hline
Source  & RA (h m s) & DEC ($^\circ$ $^{'}$ $^{''}$) & LWS01 & LWS04 & SWS02\\
\hline
 M--0.96+0.13  &17:42:48.3 &-29:41:09.1 & 31301304&31301303  &31301401, 46901501\\
 M--0.55--0.05 &17:44:31.3 &-29:25:44.6 & 46901403&46901404  &46901302\\
 M--0.50--0.03 &17:44:32.4 &-29:22:41.5 & 32201017&32201018  &31500616, 46901216\\
 M--0.42+0.01  &17:44:35.2 &-29:17:05.4 & 32702112&32702113  &32700711, 46901111\\
 M--0.32--0.19 &17:45:35.8 &-29:18:29.9 & 48501456&48501457  &48501055 \\
 M--0.15--0.07 &17:45:32.0 &-29:06:02.2 & ...     & ...      &48501659\\
 M+0.16--0.10  &17:46:24.9 &-28:51:00.0 & 48502308&48502309  &48502207 \\
 M+0.21--0.12  &17:46:34.9 &-28:49:00.0 & 49401613&49401614  &48502612\\
 M+0.24+0.02   &17:46:07.9 &-28:43:21.5 & 32701822&32701823,46900423  &31502321, 46701921\\
 M+0.35--0.06  &17:46:40.0 &-28:40:00.0 & 49401518&49401519  &49401817\\
 M+0.48+0.03   &17:46:39.9 &-28:30:29.2 & 49401430&49401431  &49401929\\
 M+0.58--0.13  &17:47:29.9 &-28:30:30.0 & 49400833&49400834  &49402032\\
 M+0.76--0.05  &17:47:36.8 &-28:18:31.1 & 31300507&31300508  &31301206, 46701706\\
 M+0.83--0.10  &17:47:57.9 &-28:16:48.5 & 50500238&50500239  &47002037\\
 M+0.94--0.36  &17:49:13.2 &-28:19:13.0 & ...     &...       &47002142 \\
 M+1.56--0.30  &17:50:26.5 &-27:45:29.7 & 31801737&31801738  &31301636, 46701636\\
 M+2.99--0.06  &17:52:47.6 &-26:24:25.3 & 31801547&31801548  &31801646, 46701546\\
 M+3.06+0.34   &17:51:26.5 &-26:08:29.2 & 31502142&31502143  &31300641, 46701441\\
 \hline
\end{tabular}
\end{table*}

\subsection{ISO SWS02 observations}

\begin{figure*}[htb]
\includegraphics[bb=67 27 589 615,width=16cm]{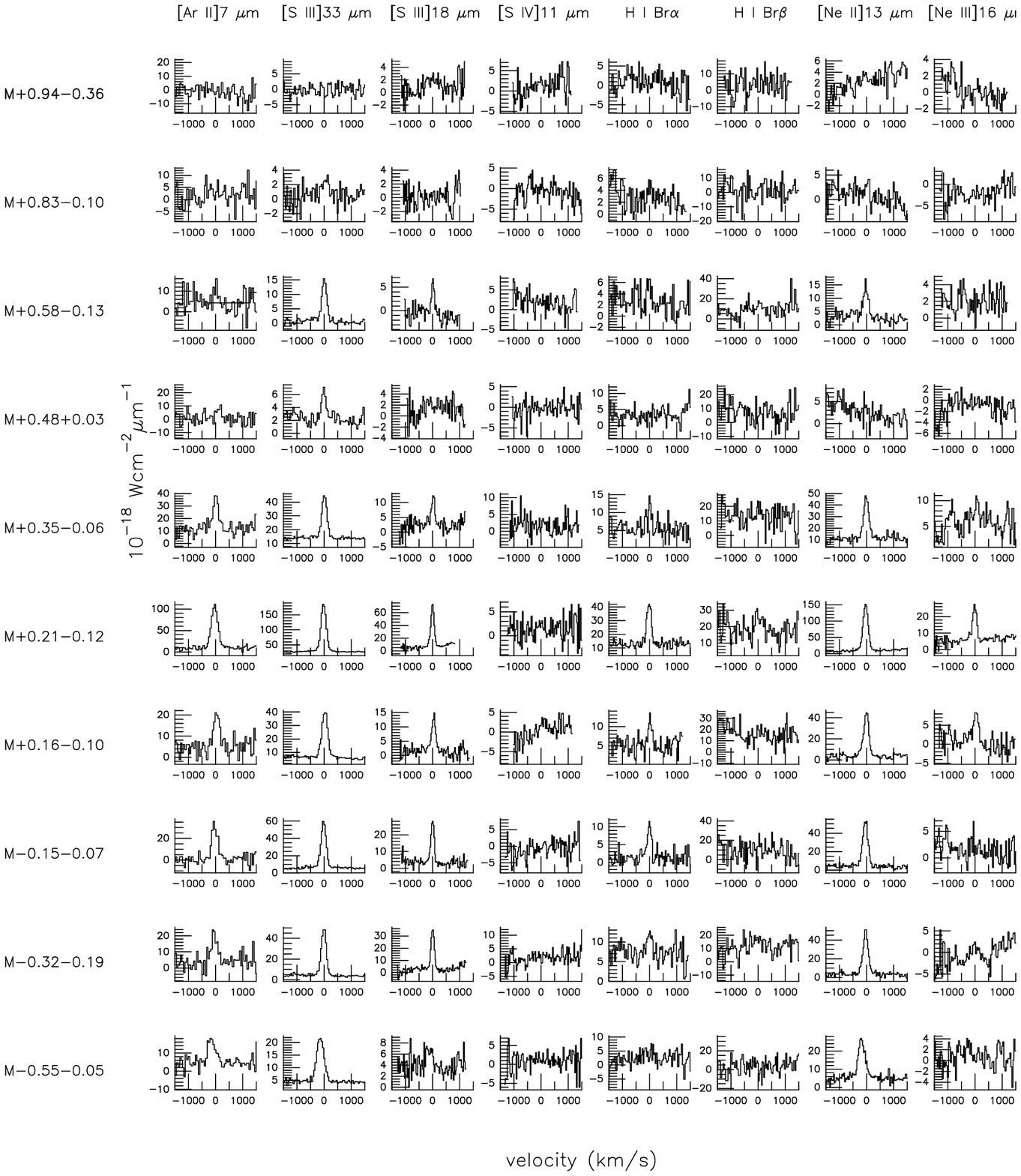}
\caption{Sample of the spectra obtained in the SWS02 mode.  Typical linewidths are $\sim 200$ \kms ~ (the instrumental spectral resolution)}
\label{fig_sws02}
\end{figure*}

We have taken spectra of individual lines in grating mode with the {\it Short Wavelength Spectrometer} (SWS02 mode) on ISO.  The Target Dedicated Time (TDT) numbers, which identify the ISO observations (in particular in the ISO Data Archive), are shown in Table \ref{tab_isoobs}. The observations were processed through the {\it Offline processing software} (OLP) version 10.0. Further data reduction was done with the {\it ISO Spectral Analysis Package} (ISAP) version 2.1. We have zapped bad data points and averaged the two scan directions for each detector. We have shifted the spectra taken with the different detectors in additive mode as recommended for medium brightness sources. Only for a few very bright lines ($>1000$ Jy) toward M+0.21-0.12 and M+0.35-0.06 was the shifting  done in gain mode. The final spectra were obtained by averaging the data of the same line taken by the different detectors. A sample of the spectra is shown in Fig. \ref{fig_sws02}. We have fitted baselines of order 0 or 1 (only order 0 when no line is detected) to the  spectra. None of the lines is resolved at this spectral resolution and the line  profiles are Gaussian, as expected from the instrumental response. We have fitted Gaussian curves to the detected lines using ISAP. The results are shown in Tab. \ref{tab_sws02}. This table also shows the wavelength of the lines as well as the telescope aperture and the velocity resolution at those wavelengths. The absolute flux calibration uncertainties of the SWS02 observing mode are less than  30 $\%$ while the wavelength calibration uncertainties at wavelength $\lambda$ are lower than $\lambda/5000$ (Leech et al. 2003).

\begin{table*}[htb]
\caption[]{Integrated flux densities of the hydrogen recombination lines and fine structure lines observed in the  SWS02 mode (in units of 10$^{-20}$ Wcm$^{-2}$).  Upper limits are the $3\sigma$ values integrated in a width equal to the spectral resolution of the SWS02 mode. Numbers in parentheses are the errors of the last significant digit}
\label{tab_sws02}
\begin{tabular}{lllllllll}
\hline
\noalign{\smallskip}
Line      &HI Br$\alpha$& HI Br$\beta$& NeII & NeIII  & SIII & SIII& SIV &ArII\\
$\lambda (\mu\mathrm{m})$ & 4.05  & 2.62  &12.81&15.55& 33.48&18.71&10.51&6.98   \\
Aper. ($^{"}\times^{"}$) & 14$\times$20& 14$\times$20&14$\times$27&14$\times$27& 20$\times$33& 14$\times$27 & 14$\times$20&  14$\times$20 \\
Res. (\kms) & 207 & 193 & 214 &176 & 214 & 143 & 133 & 250 \\
\noalign{\smallskip}
\hline
\noalign{\smallskip}
M--0.96+0.13 &\le 0.60  &\le 1.65  & 7.5(11) &\le2.2    & 13(2)    &\le1.8    &---    &\le2.4  \\
M--0.55--0.05&\le 0.73  &\le 1.3   &27.4(14) &\le2.1    & 65(2)    &\le3.0    & \le1.8&11(2)   \\
M--0.50--0.03&1.6(2)    &\le 1.6   & 22.3(9) &\le2.2    & 64(2)    &7(2)      & ---   &8.3(13) \\
M--0.42+0.01 &1.5(5)    &\le 2.1   &24.3(11) &\le2.2    & 58(3)    & 9.5(11)  &---    &10.4(11)\\
M--0.32--0.19&1.5(3)    &\le 1.2   &46(2)    &2.7(5)    & 117(3)   &30.8(7)   &\le1.5 &11(2)   \\
M--0.15--0.07&2.3(3)    &\le 1.4   & 50(2)   &\le2.3    & 141(4)   &23.4(12)  &\le1.7 &19(3)   \\
M+0.16--0.10 &1.6(3)    &\le 1.4   &41(2)    &9.3(13)   & 102(3)   &14.4(10)  &1.9(5) &10.5(14)\\
M+0.21--0.12 &8.2(6)    &1.7(6)    &139(2)   &20(2)     & 425(12)  &63(2)     &\le1.5 &69(2)   \\
M+0.24+0.02  &\le 0.59  &\le 1.7   &13.8(11) &\le2.2    & 126(4)   &3.2(6)    &---    &3.8(14) \\
M+0.35--0.06 &1.8(4)    &\le 1.1   &36.2(14) &3.0(5)    &83(3)     &9.9(2)    &\le 1.7&15.4(12)\\
M+0.48+0.03  &\le 0.77  &\le 1.2   &\le 2.1  &\le1.9    &9.6(9)    &\le1.8    & \le1.4&\le 4.5 \\
M+0.58--0.13 &0.9(2)    &\le 1.1   &11.7(2)  &\le1.5    &36(2)     & 6.4(9)   &\le 1.5&3.6(11) \\
M+0.76--0.05 &\le 0.46  &\le 2.0   &2.5(4)   &\le1.5    & 9.0(10)  &\le1.4    &---    &\le 4.6 \\
M+0.83--0.10 &\le0.5    &\le 1.0   &2.0(8)   &\le2.5    & 5.9(10)  &\le1.6    &\le 2.4&\le2.7  \\
M+0.94--0.36 &\le 0.67  &\le 0.90  &\le 1.5  &\le1.3    & \le5.8   &\le1.6    &\le 1.6&\le 4.2 \\
M+1.56--0.30 &\le 0.57  & \le 1.1  &\le 1.2  & ---      & \le 3.2  &---       &---    &---     \\
M+2.99--0.06 &\le 0.72  &\le 1.7   &3(2)     &\le1.5    & 11(2)    &2.6(6)    &---    &\le4.0  \\
M+3.06+0.34  &\le 0.57  &\le 1.4   &\le 1.4  & ---      &2.0(5)    & ---      &---    &---     \\
\hline
\end{tabular}
\end{table*}

\subsection{ISO LWS01 observations}

\begin{figure}[h]
\includegraphics[bb=68 37 425 514,width=8cm]{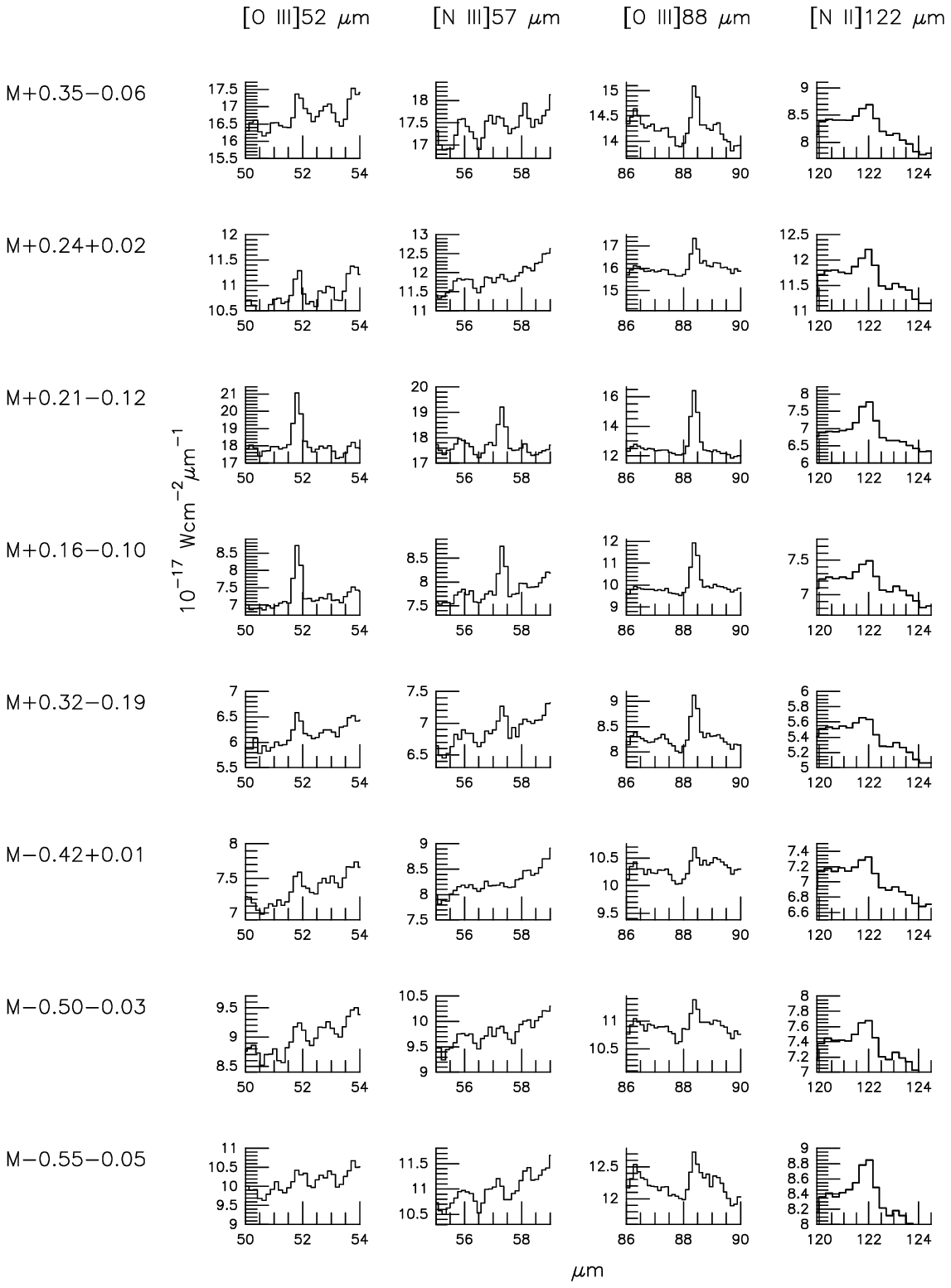}
\caption{ Sample of the LWS01 line spectra discussed in this paper. The [\CII] and [\OI] lines are discussed in Rodriguez-Fernandez et al. (2004)}
\label{fig_lws01}
\end{figure}

We have also taken continuous spectra in the 43-197 \mum \, range with the {\it Long Wavelength Spectrometer} (LWS01 mode). This wavelength range is scanned with five ``short wavelength'' (SW) detectors (43-93 \mum) and five ``long wavelength'' (LW) detectors (84-197 \mum). The spectral resolution of this mode is 0.29 and 0.6 \mum \, for the SW and LW detectors, respectively. The aperture of the LWS is $\sim 80^{''} \times 80^{''}$  (the exact values  can be found in Gry et al. 2003 and are shown in Table \ref{tab_lws01}).  Table \ref{tab_isoobs} shows the TDT numbers of the observations.  The observations were processed through the OLP software version 10.0. Further reduction was done with ISAP 2.1.We have zapped bad data points and shifted the different scans to a common level. Afterwards we have averaged the different scans. The absolute flux calibration uncertainties for the LWS01 mode are smaller than $\sim 20\%$  (Gry et al. 2003).  We have fitted order 1 baselines to the spectra in the vicinity of the lines and Gaussian curves to the lines using ISAP. A sample of the detected lines is displayed in Fig. \ref{fig_lws01} and the line fluxes are shown in Table \ref{tab_lws01}.

\begin{table}[htb]
\caption[]{Integrated flux densities of the lines observed in the LWS01 mode in units of  10$^{-19}$ Wcm$^{-2}$. Upper limits are 3$\sigma$ flux densities integrated in a velocity range equal to the velocity resolution of the LWS01 mode. Numbers in parenthesis are errors in the last significant digits.  The table also contains the line wavelengths and telescope apertures (Gry et al. 2003)}
\label{tab_lws01}
\begin{tabular}{llllll}
\hline
\noalign{\smallskip}
Line      &\OIII    & \NIII   & \OIII  & \NII   \\
 $\lambda (\mu\mathrm{m})$      &52 \mum  & 57 \mum &88 \mum & 122 \mum \\
 Aper. ($^{"}\times^{"}$) & 79.9$\times$77.3& 79.9$\times$77.3&83.5$\times$76.0&78.1$\times$74.9 \\
\noalign{\smallskip}
\hline
\noalign{\smallskip}
M--0.96+0.13  &  \le3.4 & \le 3.4 & 7(2)       &4.4(5)  \\
M--0.55--0.05 & \le 9.7 & \le 9.7 & 16(2)      & 44(3)  \\
M--0.50--0.03 & \le 8.4 & \le 8.4 & 18(2)      & 27(2)  \\
M--0.42+0.01  &\le 11   & \le 7.1 & 10(2)      & 23(2)  \\
M--0.32-0.19  & 16(2)   & 10(3)   & 28(3)      & 22(2)   \\
M--0.15--0.07 & ...     & ...     & ...        & ...    \\
M--0.16--0.10 & 50(3)   & 23(4)   & 65(5)      & 26(2)  \\
M+0.21--0.12  & 108(5)  & 48(3)   &118(4)      &74(5)   \\
M+0.24+0.02   & 14(4)   & \le 11  & 35(5)      &39(3)   \\
M+0.35--0.06  & \le 22  & \le17   & 23(6)      &31(2)   \\
M+0.48+0.03   & \le 6.5 &\le 6.5  &\le 7.5     & 17(4)  \\
M+0.58-0.13   & \le 7.0 &\le 7.0  &16(2)       & 21(3)  \\
M+0.76-0.05   & \le 5.6 &\le 5.6  &\le7.9      &\le 13  \\
M+0.83--0.10  &\le 5.6  &\le 5.6  & \le 3.9    &\le 6.5 \\
M+0.94--0.36  & ...     & ...     & ...        & ...    \\
M+1.56-0.30   & \le 3.3 & \le 3.3 & \le 1.6    & \le 1.5 \\
M+2.99--0.06  & \le 3.2 & \le 3.1 & \le 3.0    & 3.7(5)  \\
M+3.06+0.34   & \le 3.1 & \le 3.1 & \le 1.5    & \le1.5  \\
\noalign{\smallskip}
\hline
\end{tabular}
\end{table}

\subsection{ISO LWS04 observations}

\begin{figure}[htb]
\includegraphics[bb=30 37 498 648,width=8cm]{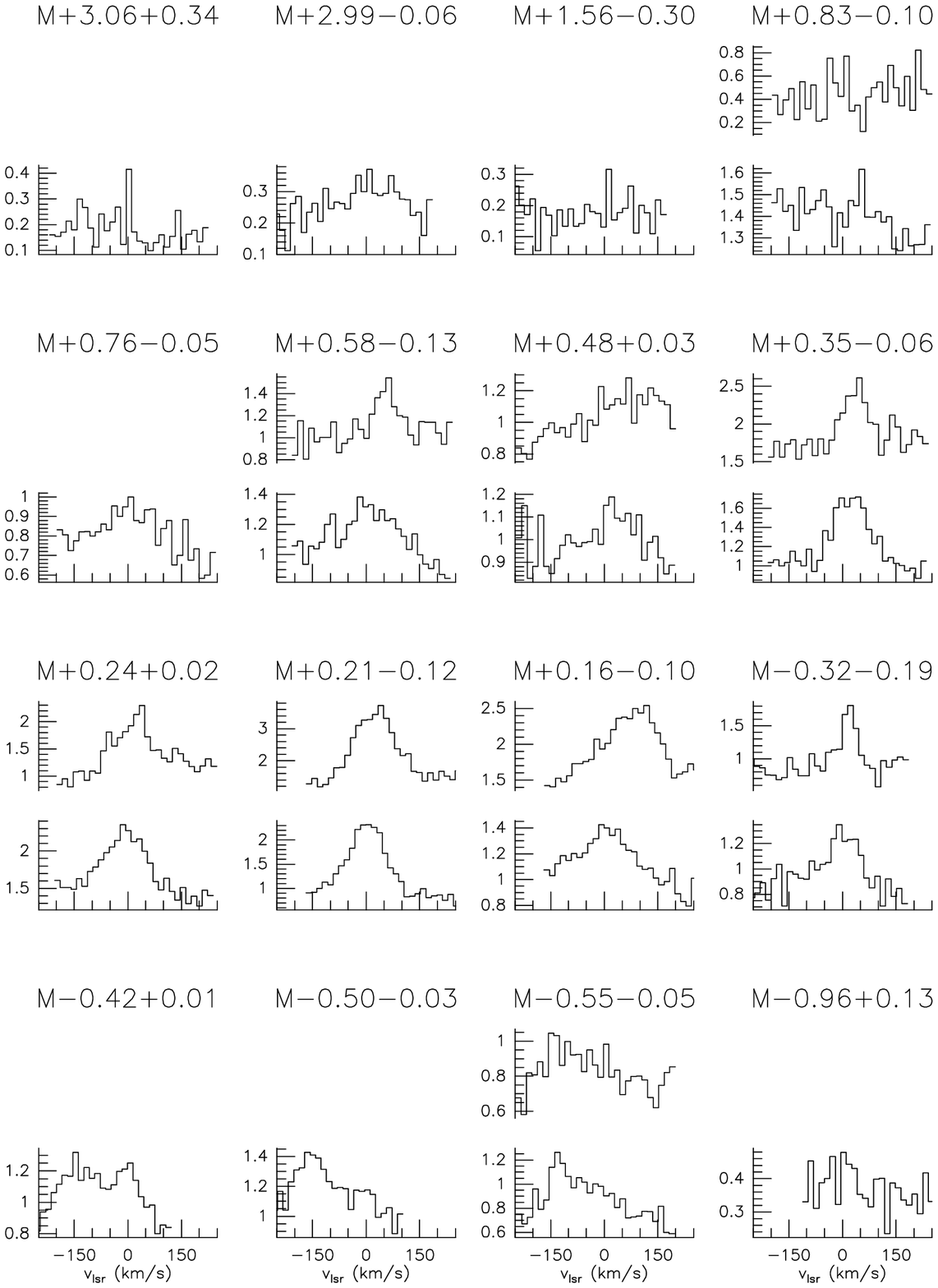}
\caption{ LWS04 spectra. For each source, the lower spectrum is the [\NII] 122 \mum~ line and the upper spectrum is the [\OIII] 88 \mum~ line}
\label{FPvsCO_all.eps}
\end{figure}

Using the LWS we have also taken spectra of individual lines with the Fabry-Perot (LWS04 mode), which gives a velocity resolution of $\sim 30$ \kms. The line flux calibration uncertainties for this mode are $\sim 30 \%$  (Gry et al. 2003).   Table \ref{tab_isoobs} shows the TDT numbers of the observations.  The observations were processed trough the OLP software 10.1 and ISAP 2.1. With ISAP we have determined the active detector for each line and we have zapped bad data points before averaging the 7 scans available for each line. A sample of the spectra is shown in Fig. \ref{FPvsCO_all.eps}. The spectra have been fitted with order 0 or 1 baselines and Gaussian curves. Fluxes, linewidths and the velocities of the center of the lines are listed in Table \ref{tab_lws04}. At the resolution of the LWS04 mode all the lines have been resolved. Most of them show large linewidths up to $\sim 150$ \kms. The central velocities and the profiles of the different lines detected in a given  source are in agreement with each other.

The LWS04 and the LWS01 line  fluxes are in good agreement. For $\sim 70 \%$ of the observed lines both fluxes differ by less than 30 $ \%$. For the rest of the lines, LWS04 and LWS01 fluxes differ by less than $50\%$. Both results are within the calibration uncertainties  (see also Gry et al. 2003).

\begin{table}[htb]
\caption[]{LWS04  integrated flux densities in units of 10$^{-18}$ W \cmmd \smu.  Linewidths in parenthesis are the values used to estimate upper limits to the  integrated flux density when the line is not detected. Other numbers in parenthesis  are the errors in the last significant digit of the flux densities.  The table  also contains the linewidth ($\Delta v$) and the line peak heliocentric velocity ($v_{hel}$) in \kms}
\label{tab_lws04}
\begin{tabular}{lllllll}
\hline
\noalign{\smallskip}
Source&\multicolumn{3}{c}{\NII~ 121.89~\mum}&
\multicolumn{3}{c}{\OIII~88.35~\mum}\\
~    &Flux & $\Delta v$ & $v_{hel}$ & Flux & $\Delta v$  &  $v_{hel}$ \\
\noalign{\smallskip}
\hline
\noalign{\smallskip}
M-0.96+0.13& \le0.3  & (100)  & &...&...&...\\
M-0.55-0.05& 0.8    & 45  & -150   &1.4(3)   & 217 & -120\\
~   &        1.9(4) & 150 &-60     & & & \\
M-0.50+0.03& 2.1(3) & 140  &-160 &...&...&...\\
M-0.42-0.01& 1.8(30)  & 120  & -150 &...&...&...\\
~          & 1.2(4) & 90 &  15&  & &  \\
M-0.32-0.19& 1.0(7)   & 70  & 0 & 1.28(6) & 42  & 9 \\
~          & 2.5(6)   & 200 & -45 &    & & \\
M-0.15-0.07&...&...&...&...&...&... \\
M+0.16-0.10& 2.3(2)   & 142  &  4 & 4.4(5)   & 150  & 69 \\
M+0.21-0.12& 7.0(3)   & 113 & -6   & 8.9(5)   & 130 & 13  \\
M+0.24+0.02&4.2(3) & 120   & 16 & ...&... &... \\
~          & ...      &  ... & ...   &3.4(7)&125&48 \\
M+0.35-0.05&3.5(4)   & 112   & 14  & 1.7(3)   & 73    & 24 \\
M+0.48+0.03& 0.8(2)  & 100     & -14   & \le 0.4  & (90)   &    \\
M+0.58-0.13& 2.1(2)  & 120 & -27 & 0.90(14) & 63 & 46  \\
M+0.76-0.05&\le0.7  & (140)  &  &...&...&... \\
M+0.83-0.10& \le 0.6  & (140)     &    & \le 0.7  & (140) &  \\
M+0.94-0.36&...&...&...&...&...&...\\
M+1.56-0.30& \le0.3  & (150)  & &...&...&... \\

M+2.99-0.06& \le0.3  & (100)  & &...&...&...\\
M+3.06+0.34& \le0.3  & (100)  &  &...&...&...\\
\noalign{\smallskip}
\hline
\end{tabular}
\end{table}

\subsection{IRAM-30m recombination lines observations}

 We have obtained upper limits to the intensity of several hydrogen radio recombination lines with the IRAM 30 meter radio telescope. The observations were carried out in May  1997 and September 1998. We have observed the H35$\alpha$, H39$\alpha$ and H41$\alpha$ lines. The frequencies of the lines, the telescope beam and the velocity resolution of the 1 MHz filter banks at those frequencies are listed in Table \ref{tab_reclin}. We have not detected any of the lines.  Table \ref{tab_reclin} shows the upper limits to the intensity of the lines. The quoted upper limits are the 3$\sigma$ dispersion of the spectra obtained by fitting order 1 baselines. The intensity calibration uncertainties are  $\sim 10 \%$ at 3 mm (H39$\alpha$, H41$\alpha$) and $<20\%$ at 2 mm (H35$\alpha$; see for  instance, Mauersberger et al. 1989).  The upper limits to the intensities have been used to constrain the Lyman continuum photons emission rate (see Sect. \ref{sect_lyc}).

\begin{table}[htb]
\caption[]{ Upper limits (3$\sigma$) to the antenna temperature, $T_A^*$ (in units of  K) of the radio recombination lines and upper limits to the Lyman continuum photons emission rate, $Q$(H) (in units of \smu), derived from the recombination line data}
\label{tab_reclin}
\begin{tabular}{lllllll}
\hline
\noalign{\smallskip}

Line  & \multicolumn{2}{c}{H35$\alpha$} &
   \multicolumn{2}{c}{H39$\alpha$} & \multicolumn{2}{c}{H41$\alpha$} \\
Freq.(MHz) & \multicolumn{2}{c}{147046.901} & \multicolumn{2}{c}{106737.374}&
 \multicolumn{2}{c}{92034.449}\\
Beam (arcsec) & \multicolumn{2}{c}{16} &  \multicolumn{2}{c}{23} &
 \multicolumn{2}{c}{27}\\
Res. (\kms) & \multicolumn{2}{c}{2.0} & \multicolumn{2}{c}{3.5} &
\multicolumn{2}{c}{3.3} \\
  & $T_A^*$ & $Q$(H)  &  $T_A^*$ & $Q$(H) & $T_A^*$ & $Q$(H) \\
\noalign{\smallskip}
\hline
\noalign{\smallskip}
M-0.96+0.13 & \le  0.10 &\le47.8 &      ...  & ...     & \le 0.07 &\le48.0 \\
M-0.55-0.05 & \le  0.05 &\le47.5 & \le  0.04 &\le47.7  & \le 0.05 &\le47.9 \\
M-0.50-0.03 & \le  0.3  &\le48.4 &      ...  & ...      & \le 0.08 &\le48.1 \\
M-0.42+0.01 & \le  0.3  &\le48.2 &      ...  & ...      & \le 0.07 &\le48.0 \\
M-0.32-0.19 & \le  0.3  &\le48.2 & \le  0.05 &\le47.8 & \le 0.03 &\le47.7 \\
M-0.15-0.07 &      ...  &    ... & \le  0.04 &\le47.7 & \le 0.04 &\le47.8 \\
M+0.16-0.10 &     ...   &    ... & \le  0.04 &\le47.6 & \le 0.02 &\le47.6 \\
M+0.21-0.12 & \le  0.07 &\le47.7 & \le  0.06 &\le47.9 & \le 0.03 &\le47.7 \\
M+0.24+0.02 & \le  0.05 &\le47.5 & \le  0.04 &\le47.7 & \le 0.04 &\le47.8 \\
M+0.35-0.06 & \le  0.3  &\le48.2 & \le  0.04 &\le47.7 & \le 0.10 &\le48.2 \\
M+0.48+0.03 & \le  0.05 &\le47.5 &      ...  & ...      & \le 0.04 &\le47.7 \\
M+0.58-0.13 & \le  0.05 &\le47.6 &      ...  & ...      & \le 0.07 &\le48.0 \\
M+0.76-0.05 & \le  0.04 &\le47.5 & \le  0.04 &\le47.8 & \le 0.04 &\le47.8 \\
M+0.83-0.10 & \le  0.2  &\le48.1 & \le  0.10 &\le48.1 &     ...   &    ... \\
M+0.94-0.36 & \le  0.2  &\le48.1 &   ...     & ...      & \le 0.09 &\le48.1 \\
M+1.56-0.30 & \le  0.5  &\le48.5 &   ...     & ...      & \le 0.08 &\le48.1 \\
M+2.99-0.06 & \le  0.2  &\le48.1 &   ...     & ...      & \le 0.09 &\le48.2 \\
M+3.06+0.34 & \le  0.2  &\le48.1 &   ...     & ...      & \le 0.08 &\le48.1 \\
\noalign{\smallskip}
\hline
\end{tabular}
\end{table}

\section{Extinction correction}

The average visual extinction toward the Galactic center is $\sim 25$ magnitudes. Therefore, extinction is important even at mid infrared wavelengths  and the ISO line fluxes must be corrected for this effect. Since our sources are molecular peaks located in the GC, the average foreground extinction derived from star counts can be considered as a lower limit to the actual extinction of the lines arising from these GC sources.  By comparing with the extinction maps from Catchpole et al. (1990) and Schultheis et al. (1999) one finds that the visual extinction obtained from star counts is $A_V =12-15$ mag for the Clump 2 and the l=1.3$^{\circ}$-{\it complex} sources, and range from $A_V \sim 20$ for the $l\sim \pm 1^\circ$ sources to $A_V \sim 30$ for the sources located close to the center of the Galaxy. We have also used our [\SIII] data to estimate a lower limit to the extinction. In most of the sources the observed [\SIII] 18 to 33 \mum \ line ratio is lower than the lower limit ($\sim 0.5 $) for very low electron density. This is because the extinction at 18 \mum \ is higher than that at 33 \mum. We have calculated the minimum visual extinction necessary to have a [\SIII] 18 to 33 \mum \ ratio of 0.5 assuming the extinction law  of Draine (1989) or Lutz (1999) and that the emission is extended (to correct for the slightly different beam sizes at the frequencies of those lines). The minimum extinction derived with this method varies from $\sim 3$ mag for M+0.83-0.10 and M-0.32-0.19 to 60-80 mag for M-0.55-0.05 and M+0.24+0.02. For most of the sources the minimum extinction derived from the [\SIII] data  is 20-30 mag, in agreement with the average foreground extinction derived from star counts. Only for M-0.55-0.05 and M+0.24+0.02 do we find direct  evidence of an extinction {\it intrinsic} to our sources (a minimum extinction much higher than the average foreground extinction). The lower limits to the visual extinction  that we have assumed throughout this work are shown in Table \ref{tab_Av}. These have been derived from the [\SIII] \ lines for all the sources except for those where the [\SIII] \ lines were not detected or the derived lower limit was much lower than the foreground extinction. For these sources we have taken a lower limit to the visual extinction of 15 mag.

To obtain upper limits to the extinction we have compared infrared and radio hydrogen recombination lines. From the upper limits to the radio lines fluxes we have calculated upper limits to the emission measure for each source and with these, we have estimated upper  limits to the fluxes of the infrared lines. Comparing these upper limits with the observed Br$\alpha$ fluxes one gets an upper limit to the extinction that  varies from source to source between 50 and 80 mag. For a given source these upper limits are always higher than  the lower limits derived from the [\SIII] \ lines. Unfortunately, the former method is only applicable to the eight sources with detected Br$\alpha$ and it is not exempt of uncertainties. Therefore, we have also estimated  upper limits to the extinction from the column density of molecular gas in the clouds. We have transformed the H$_2$ column densities measured by Rodríguez-Fernández et al. (2000, 2001b) in visual extinction following the equation $A_V (mag) = 10^{-21} \times N_{H_2}$(\cmmd)  (Bohlin 1975). In order to obtain a conservative upper limit to the extinction we have added these values to the average foreground extinction as derived from star counts. The upper limits derived in this way are listed in Table \ref{tab_Av}. These values are comparable or somewhat higher than those derived from the hydrogen recombination lines for the sources where the Br$\alpha$ was detected.

The analysis of the infrared lines in the following has been done for both the minimum and maximum extinction corrections. To get the actual extinction at a given mid-infrared wavelength we have used Draine (1989) and Lutz (1999) extinction laws. The results obtained using these  laws are similar for all but the [\ArII] 6.98 \mum~ line since it is at the wavelength of this line that these  extinction laws differ the most. In the following we use the extinction law derived by Lutz. [\ArII] 6.98 \mum~ fluxes would be a factor of $\sim 2-4$ lower if the mid-infrared extinction law has  the 7 \mum \, minimum predicted by Draine (1989). For the extinction in the LWS range we have taken a 40 to 0.55 \mum ~ (A$_{40}$/A$_V$) extinction ratio of 0.01 from Lutz  (1999) and we have extrapolated to larger wavelengths using a power law with an exponent of 1.5.

\begin{table}
\caption{Upper and lower limits to the extinction toward the sources studied in this paper}
\label{tab_Av}
\begin{tabular}{llll}
\hline
 Source  & Av   & Source & Av\\
 \hline
M-0.96+0.13 &  24- 35 & M+0.35-0.06 &  30- 55  \\
M-0.55-0.05 &  60- 90 & M+0.48+0.03 &  14- 55  \\
M-0.50-0.03 &  32- 60 & M+0.58-0.13 &  16- 65  \\
M-0.42+0.01 &  18- 55 & M+0.76-0.05 &  20-110  \\
M-0.32-0.19 &  15- 45 & M+0.83-0.10 &  15- 85  \\
M-0.15-0.07 &  18-105 & M+0.94-0.36 &  15- 50  \\
M+0.16-0.10 &  23- 75 & M+1.56-0.30 &  15- 35  \\
M+0.21-0.12 &  22- 45 & M+2.99-0.06 &  15- 34  \\
M+0.24+0.02 &  80- 95 & M+3.06+0.34 &  15- 29  \\
\hline
\end{tabular}
\end{table}

\section{Lyman continuum photons emission rate}
\label{sect_lyc}

 The upper limits to the radio recombination lines intensities obtained in Sect. 2 can be used to derive significant upper limits to the Lyman continuum photons emission rate, $Q(H)$, via the equation:

\begin{equation}
Q(H)=2.81\,10^{38} \, \nu \, {T_e}^{0.7} \,  D^2 \, \theta^2
\, \frac{T_L}{b_n} \, \Delta \mathrm{v}
\label{QH}
\end{equation}

where ${T_e}$ is the electron temperature in K, $\nu$ is the frequency of the line in GHz, $\theta$ is the telescope beam expressed in arcsec (and assumed to be Gaussian), $D$ is the distance to the source in kpc, $T_L$ is the line temperature in K, $\Delta \mathrm{v}$ the line width in \kms \ and $b_n$ is the departure coefficient from the Saha-Boltzman populations for level $n$.

To obtain upper limits to $Q(H)$ in the GC sources we have taken the frequencies and the telescope beam for each line, a $b_n$ of 0.7 (Walmsley 1990), a distance $D$ of 8.5 kpc, and electron temperatures  typical of \HII ~ regions in the inner Galaxy of 6300 K  (Martín-Hernandez et al. 2002). Since none of the lines have been detected in any source we have derived an upper limit to the $T_L \Delta \mathrm{v}$ product as the $3\sigma$ dispersion in a velocity range of 200 \kms~\ (to obtain conservative limits). These upper limits to $Q(H)$ are listed in Table \ref{tab_reclin}.

On the other hand, Table \ref{qh_ir} shows the Lyman continuum photons emission rate derived from the two hydrogen recombination lines observed in the infrared.
These have been calculated in the Case B approximation  (which consist of  assuming that all the lines but those of the Lyman series are optically thin, see Storey \& Hummer 1995)  from the following expressions:

\begin{equation}
Q(H)_{Br\alpha}=4.41\,10^{57} T_4^{0.1} D^2 F_\mathrm{Br\alpha}(\mathrm{erg}\,\mathrm{s}^{-1}\,\mathrm{cm}^{-2})
\label{ccc}
\end{equation}

\begin{equation}
Q(H)_{Br\beta}=6.86\,10^{57} T_4^{0.1} D^2 F_\mathrm{Br\beta}(\mathrm{erg}\,\mathrm{s}^{-1}\,\mathrm{cm}^{-2})
\end{equation}

where $T_4$ is the electron temperature in units of 10$^4$ K, $D$ the distance
to the source in kpc and $F$ the integrated flux of the line in units of erg\,s$^{-1}$\,cm$^{-2}$. We have taken the same electron temperature and distance that we took for the radio lines. For each source we give two estimates  of $Q(H)$ corresponding to the low and high extinction corrections. For the sources where Br$\alpha$ has been detected, $Q(H)_{Br\alpha}$ is, in general, lower than the upper limits derived from radio recombination lines. The exceptions are the $Q(H)_{Br\alpha}$ derived for M-0.15-0.07 and M+0.16-0.10 in the upper limit to the extinction and that derived for M+0.21-0.12 both in the low and high extinction limits. The discrepancy between radio and the infrared determinations of $Q(H)$ in the high extinction limit implies that the actual extinction of the infrared lines is lower than the upper limit of Table \ref{tab_Av}. This fact is in agreement with the extinction correction used by Rodríguez-Fernández et al. (2001a) for the Radio Arc sources ($A_V=30$). The discrepancy between the $Q(H)$ derived for M+0.21-0.12 from Br$\alpha$ in the low extinction limit and the upper limits derived with H35$\alpha$ and H41$\alpha$ is very low (0.2 dex). Indeed, both results are in agreement taking into account the calibration uncertainties and the slightly different beam sizes.

The measured values of $Q(H)_{Br\alpha}$, $10^{47}-10^{48}$ \smu, are equivalent to those produced by stars with a spectral type later B0 and an effective temperature lower than $\sim$ 32000 K (Schaerer \& de Koter 1997). 

\begin{table}
\caption{Lyman continuum photons emission rate (\smu) as derived from
the Br$\alpha$ and Br$\beta$ lines.}
\label{qh_ir}
\begin{tabular}{lll}
\hline
 Source     & Q(H)Br$\alpha$   & Q(H)Br$\beta$  \\
 \hline
M-0.96+0.13 &   \le46.8-47.0 &  \le47.6-48.0 \\
M-0.55-0.05 &   \le47.6-48.3 &  \le48.6-49.5 \\
M-0.50-0.03 &      47.3-47.9 &  \le47.9-48.7 \\
M-0.42+0.01 &      47.1-47.9 &  \le47.6-48.7 \\
M-0.32-0.19 &      47.0-47.6 &  \le47.2-48.1 \\
M-0.15-0.07 &      47.2-49.1 &  \le47.4-50.0 \\
M+0.16-0.10 &      47.2-48.3 &  \le47.5-49.1 \\
M+0.21-0.12 &      47.9-48.4 &     47.6-48.3 \\
M+0.24+0.02 &   \le48.0-48.3 &  \le49.3-49.8 \\
M+0.35-0.06 &      47.4-47.9 &  \le47.6-48.4 \\
M+0.48+0.03 &   \le46.7-47.6 &  \le47.2-48.4 \\
M+0.58-0.13 &      46.8-47.9 &  \le47.2-48.7 \\
M+0.76-0.05 &   \le46.6-48.6 &  \le47.6-50.3 \\
M+0.83-0.10 &   \le46.5-48.0 &  \le47.1-49.3 \\
M+0.94-0.36 &   \le46.7-47.4 &  \le47.1-48.2 \\
M+1.56-0.30 &   \le46.6-47.0 &  \le47.2-47.8 \\
M+2.99-0.06 &   \le46.7-47.1 &  \le47.4-47.9 \\
M+3.06+0.34 &   \le46.6-46.9 &  \le47.3-47.7 \\
\hline
\end{tabular}
\end{table}

\section{Electron densities}

The ratio of two lines from the same  fine structure ladder can be used to estimate the electron density since it is almost independent of the electron temperature. Here we  have used the [\OIII] ~ 52/88 and the  [\SIII] ~ 18/33 ratios. The [\OIII] lines  ratio  probe densities lower than a few thousand particles per cubic centimeter while the [\SIII] lines ratio probes higher density gas due to the higher critical densities of these lines.

The extinction corrected [\OIII] ~ 52/88 ratio derived from our observations varies between $\sim 1$ for the sources located at low Galactic longitudes to $\leq 0.6$ for the sources located far from the center of the Galaxy (Table \ref{tab_rat1}). Comparing these [\OIII]~ ratios with  the model calculations by Alexander et al. (1999) one finds electron densities ranging from $\sim 200$ cm$^{-3}$ to $\leq 30$ cm$^{-3}$ for the sources located close and far from the center of the Galaxy, respectively (Table \ref{tab_rat1}).

On the other hand, we have also compared the extinction corrected [\SIII]  18/33 ratio (Table \ref{tab_rat1}) with the calculations by Alexander et al. (1999). However, since for most of the sources we have used this ratio to determine lower limits to the extinction, for these sources we can only determine upper limits to the electron density corresponding to upper limits to the extinction. These upper limits are typically of a few 10$^3$ cm$^{-3}$ (Table \ref{tab_rat1}). For the sources where a lower limit to the extinction has been derived from star counts, one obtains densities of $\sim200-3000$ \cmmt.

In conclusion, the electron densities derived from the [\OIII] lines for the sources in the Radio Arc  (M+0.16-0.10 and M+0.21-0.12) are comparable to  those derived with the same method for other sources in the Arc (Rodriguez-Fernandez et al. 2001a,  100-300 \cmmt) and for the Sgr B2 envelope (Goicoechea et al. 2004,  $\sim$250 \cmmt ) while sources at negative Galactic longitudes have lower densities. For the sources in the Clump 2, l=1.3$^\circ$, and SgrB2 complexes, the [\OIII] ~ lines have not been detected, however the densities derived from the [\SIII] ~ lines are similar to those derived for the rest of the sources. For comparison, we also mention that the electron densities derived with ISO for the sample of starburst galaxies of Verma et al. (2003) are 10-600 \cmmt ~ and for the sample of normal galaxies of Malhotra et al. (2001) are  10-100 \cmmt.

\begin{table}[tbh]
\caption{Fine structure lines ratios sensitive to the electron density
and densities derived with those ratios}
\label{tab_rat1}
\begin{tabular}{lllll}
\hline
\noalign{\smallskip}
Source   & [\OIII] 52/88 & n$_e$\cmmt &[\SIII] 18/33 & n$_e$\cmmt  \\
\noalign{\smallskip}
\hline
\noalign{\smallskip}
M-0.96+0.13  &\le 0.55  &\le 25  & \le0.8  & \le400   \\
M-0.55-0.05  &\le 0.84  &\le 80  & \le1.6  & \le1000   \\
M-0.50-0.03  &\le 0.58  &\le 29  & 0.5-1.4 & \le1000   \\
M-0.42+0.01  &\le 1.3   &\le 30  & 0.5-1.8 & \le2000   \\
M-0.32-0.19  &0.60-0.67 &32-44   & 0.8-2.0 & 500-2000 \\
M-0.15-0.07  &...       &...     & 0.5-9.0 &\le10$^4$      \\
M+0.16-0.10  &0.84-1.0  &80-140  & 0.5-2.9 & \le500   \\
M+0.21-0.12  &0.99-1.1  &130-160 & 0.5-1.1 & \le800  \\
M+0.24+0.02  &0.54-0.57 &23-27   & 0.5-1.0 & \le400    \\
M+0.35-0.06  &\le 1.2   &\le 200 & 0.5-1.3 & \le1000   \\
M+0.48+0.03  &...       &...     & \le 2.0 & \le2000  \\
M+0.58-0.13  &\le 0.55  &\le 250 & 0.5-2.7 & \le3000  \\
M+0.76-0.05  &...       &...     & \le 10.0& \le10$^4$     \\
M+0.83-0.10  &...       &...     & \le7.8  &400-10$^4$    \\
M+0.94-0.36  &...       &...     & ...     &400-3000      \\
M+1.56-0.30  &...       &...     & ...     &...          \\
M+2.99-0.06  &\le 1.3   &\le270  & 0.7-1.3 &200-2000  \\
M+3.06+0.34  &...       &...     & ...     & ...           \\
\noalign{\smallskip}
\hline
\end{tabular}
\end{table}

\section{Abundances}
\label{sect_abun}
We have estimated the ionic abundances from the line fluxes assuming that the nebulae are homogeneous with constant electron temperature and density. We have also assumed that all the line photons emitted in the nebula can escape without absorption and thus no radiative transport calculations are needed. Under these assumptions the ratio of the abundances of  ion $X^{+i}$ to H$^{+}$ can be derived from the fine structure lines fluxes of ion $X^{+i}$ ($F_{X^{+i}}$) and for instance an infrared hydrogen recombination line like Br$\alpha$ ($F_{Br\alpha}$) as given by the following expression:

\begin{equation}
\frac{X^{+i}}{H^+}=\frac{F_{X^{+i}}}{F_{Br\alpha}}
                   \frac{\epsilon_{Br\alpha}}{\epsilon_{X^{+i}}}
                   \frac{\Omega_{Br\alpha}}{\Omega_{X^{+i}}}
\end{equation}

where $\epsilon_{X^{+i}}$ and $\epsilon_{Br\alpha}$ are the emissivities of the respective lines.  To take into account that the beam sizes are, in general, different, we have assumed that the emission is extended and homogeneous. Hence, one can correct the measured flux ratio by the beam ratio  ($\frac{\Omega_{Br\alpha}}{\Omega_{X^{+i}}}$). The beam ratio is close to 1 for fine structure lines observed with the SWS but it is $\sim 1/20$ for fine structure lines observed with the LWS. Therefore, in the extreme case where all the ionized gas emission arises within the smaller SWS beam, the abundance of any ion  observed with the LWS would be underestimated  by a factor of 20.

The Br$\alpha$ emissivity has been taken from   Storey \& Hummer (1995). To determine the fine structure lines emissivities we have followed the  analysis of Verma et al. (2003). For fine structure lines in the ground electronic level and in the low density limit (which is justified since the derived electron densities are lower than the critical densities of the lines) the emissivity of a line can by expressed as a function of the collision strength and the electron temperature.
The collision strengths were obtained from the IRON project (Hummer et al. 1993) and the electron temperature was considered to be 6300 K (typical of \HII ~ regions in the inner Galaxy, see for instance Martin-Hernandez et al. 2002). The ionic abundances derived for the GC sources are shown in Tables \ref{abun} and \ref{abun3}.

To estimate the element abundances we have added  the abundances of each ion of a given element and we have applied an ionization correction factor (ICF) to take into account the contribution from unobserved species. These ICFs have been calculated by Martin-Hernandez et al. (2002) for Galactic \HII ~ regions as a function of the excitation given by the Ne$^{3+}$/Ne$^{++}$ ratio. The Ne$^{3+}$/Ne$^{++}$ ratio measured in the GC nebulae, (1.2-5.6)\,10$^{-2}$, implies ICFs of $\sim 1.2$ for sulfur and $\sim 1$ for neon and nitrogen.  In contrast, the oxygen ICF can be very large and uncertain (25-100) because O$^+$ has no fine structure lines in the ground electronic level. Therefore, Table \ref{abun} only shows the O$^{++}$ abundances, which are 1-10\,10$^{-5}$. Neon abundances in the GC sources range from 2\,10$^{-4}$ to 4\,10$^{-4}$ (Table \ref{abun3}). The derived sulfur abundances are 8-30\,10$^{-6}$ (Table \ref{abun3}). Nitrogen abundances vary from source to source from $\lsim 0.2\,10^{-4}$ to $\sim 1.2\,10^{-4}$ (Table \ref{abun}). We cannot  estimate Ar abundances since we have not observed any Ar$^{++}$ line and the Ar$^{++}$ abundance is comparable to that of Ar$^{+}$ (see Verma et al. 2003, for instance). Ar$^+$ abundances are 3-9\,10$^{-6}$ (Table \ref{abun3}).

\begin{table}[tbh]
\caption{Oxygen and nitrogen abundances}
\label{abun}
\begin{tabular}{llllll}
\hline
  Source  & O$^{++}$/H$^{+}$& N$^{+}$/H$^{+}$ & N$^{++}$/H$^{+}$ & [N/H] \\
            & 10$^{-5}$     &10$^{-5}$        &10$^{-6}$     &  10$^{-5}$      \\
\hline
M-0.50-0.03  &    \ls1.8    & 1.8-7.0  & \le 1.7-5.8 &  \lsim2.0-7.6   \\
M-0.42+0.01  &    \ls2.5    & 1.6-9.4  & \le 1.4-7.3 &  \lsim1.7-10.1  \\
M-0.32-0.19  &    1.4-5.4   & 2.5-10.4 &  3.1-11.7   &  2.8-11.5       \\
M+0.16-0.10  &    0.94-9.7  & 0.64-7.8 &  1.8-17.7   &  0.82-7.97      \\
M+0.21-0.12  &    1.4-4.0   & 1.5-4.6  &  2.7-7.5    &  1.8-5.3        \\
M+0.35-0.06  &    \ls 2.6   & 1.8-5.9  & \le2.8-8.5  &  \lsim2.1-6.8   \\
M+0.58-0.13  &    \ls 4.3   & 0.15-1.6 & \le1.5-13.1 &  \lsim0.3-1.9   \\
\hline
\end{tabular}
\end{table}

\begin{table*}[htb]
\caption{Sulfur, neon and argon abundances}
\label{abun3}
\begin{tabular}{llllllll}
\hline
  Source&  S$^{++}$/H$^{+}$ & S$^{+3}$/H$^{+}$ & [S/H]  & Ne$^{+}$/H$^{+}$  &Ne$^{++}$/H$^{+}$ & [Ne/H] &Ar$^{+}$/H$^{+}$\\
          & 10$^{-6}$     &   10$^{-6}$  &  10$^{-6}$    & 10$^{-4}$          & 10$^{-6}$ &10$^{-4}$   & 10$^{-6}$  \\
\hline
M-0.50-0.03  &6.4-9.4   & ...    & \ge7.7    &2.57 & \le4.3-6.4 &$\sim$2.6  & 5.1-6.4 \\
M-0.42+0.01  &6.5-10.5  & ...    & \ge7.8    &2.24 & \le3.7-6.2 &$\sim$2.2  & 5.3-7.1 \\
M-0.32-0.19  &20.5-28.5 & \le1.03& 24.6-35.4 &4.24 &    5.2-8.0 &$\sim$4.3  & 6.1-7.7 \\
M-0.15-0.07  &8.3-16.8  & \le4.61& 10.0-25.7 &3.01 & \le1.2-4.2 &$\sim$3.0  & 4.2-8.5 \\
M+0.16-0.10  &8.3-14.8  & \le3.01& 10.0-21.4 &3.55 & 10.1-20.3  &3.7-3.8    & 4.3-6.5 \\
M+0.21-0.12  &9.0-12.6  & \le0.89& 10.8-16.2 &2.35 &    7.1-9.8 &2.4-2.5    & 7.0-8.4 \\
M+0.35-0.06  &6.2-8.8   & \le1.32&  7.4-12.1 &2.78 &    4.2-5.9 &$\sim$2.8  & 6.6-8.0 \\
M+0.58-0.13  &6.7-11.7  & \le3.13&  8.0-17.8 &1.80 & \le3.6-7.3 &$\sim$1.8  & 2.8-4.2 \\
\hline
\end{tabular}
\end{table*}

\subsection{Comparison with other sources}

Martín-Hernández et al. (2002) have studied the elemental abundances across the Galaxy by analyzing ISO observations of compact \HII ~  regions. Their source sample includes 33 sources located at different distances from the center of the Galaxy (from 3 to 14.8 kpc) and one source in the GC (Sgr C). They have found nitrogen abundances of 2-10 $10^{-4}$ for sources in the 5-kpc ring and Sgr C (hereafter inner Galaxy) and 3-10 $10^{-5}$ for sources located at more than 5 kpc from the GC (the solar abundance is $\sim 9 10^{-5}$, Grevesse \& Sauval 1998). Simpson et al. (1995) have also measured suprasolar N abundances of $\sim 3\,10^{-4}$ in the GC \HII~ regions G0.095+0.012 and G1.13-0.11. In contrast, the nitrogen abundance that we have measured   for some sources is somewhat lower ($\lsim 0.2\,10^{-4}-1.8\,10^{-4}$; Table \ref{abun}). As discussed above, the N abundances derived in this paper  can be underestimated by the differences of the LWS beam used to observe the N lines and the SWS beam used to observe the Br$\alpha$ line.

Martín-Hernández et al. (2002)  have derived a sulfur  abundance of 0.8-1.6 $10^{-5}$ for sources in the inner Galaxy, below  the solar abundance of 2 $10^{-5}$ (Grevesse \& Sauval 1998). The sulfur abundance measured by Verma et al. (2003) in starburst galaxies (0.1-1.2 $10^{-5}$) is also sub-solar. One possibility which would  explain the low interstellar sulfur abundance is depletion (see Verma et al. 2003). We have measured sulfur abundances of (0.7-3.5) $10^{-5}$ for the GC sources, in agreement with those derived by Simpson et al. (1995) for other GC \HII ~ regions (G0.095+0.012 and G1.13-0.11). The gas phase sulfur abundance in the GC is the highest in the Galaxy. One explanation is that the sulfur depletion in the GC is lower  than that in other regions of the Galaxy. This is consistent with the large abundance of refractory molecules like SiO and alcohols like ethanol, suggesting that sulfur atoms, like these molecules, have been ejected into gas phase from the grains (Martín-Pintado et al. 1997, 2000, 2001).

Martín-Hernández et al. (2002) have found neon abundances of $(2-3) 10^{-4}$  for sources in the inner Galaxy.  The abundance decreases with increasing distance to the GC to solar values (1.2  10$^{-4}$, Grevesse \& Sauval 1998) at the solar radius or even lower values at larger distances. We have measured neon abundances of 1.8-4.3 10$^{-4}$ in the GC sources. These values are similar to those found by Martín-Hernández et al. (2002) in the 5-kpc ring and Sgr C. They are also similar to the neon abundance (0.7-5\,10$^{-4}$) measured by Verma et al. (2003) in their sample of starburst galaxies observed with the SWS.

\section{Ionization state and the properties of the ionizing radiation}
\label{sect_cloudy}

Table \ref{tab_rat4} shows some of the extinction corrected ratios that depend on the effective temperature of the ionizing radiation. The most useful ratios are [\NIII]57/[\NII]122, [\NeIII]15/[\NeII]13 and  [\SIV]10/[\SIII]33 since they are independent of element abundances and telescope beam differences.  Unfortunately, we can only derive non-significant upper limits to the  [\SIV]10/[\SIII]33 ratio and it will not be further discussed. For all but a few sources, only (significant) upper limits can be derived for the [\NIII]57/[\NII]122 and [\NeIII]15/[\NeII]13 ratios. Thus, we have also used the  [\OIII]88/[\NII]122  and [\OIII]88/[\SIII]33  ratios. The [\OIII]88 and [\NII]122 lines  have been observed with the same beam but their ratio could depend not only on excitation but also on the relative oxygen and nitrogen abundances. On the other hand, the [\OIII]/[\SIII]  ratio is weakly dependent on the O and S abundances since the relative S/O abundance  is expected to be constant (Rubin et al. 1994) but it could depend on different filling factors due to the different sizes of the LWS and SWS beams. Bearing in mind that it could be a crude approximation, to account for the differences in the beam sizes we have assumed that the emission is extended and homogeneous.

\begin{table}[htb]
\caption{Some interesting fine structure line ratios sensitive to the excitation conditions}
\label{tab_rat4}
\begin{tabular}{lllll}
\hline
\noalign{\smallskip}
Source & $\frac{[\NIII]57}{[\NII]122}$ & $\frac{[\OIII]88}{[\NII]122}$ &
 $\frac{[\NeIII]15}{[\NeII]13}$  & $\frac{[\OIII]88}{[\SIII]33}$\\
\noalign{\smallskip}
\hline
\noalign{\smallskip}
M-0.96+0.13 & \le 0.88  & 1.5    &\le 0.21&   0.34-0.40\\
M-0.55-0.05 & \le 0.31  & 0.36   &\le 0.03&   0.07-0.11\\
M-0.50-0.03 & \le 0.39  & 0.64   &\le 0.06&   0.13-0.19 \\
M-0.42+0.01 & \le 0.38  & 0.41   &\le 0.07&   0.08-0.14  \\
M-0.32-0.19 & 0.48-0.54 & 1.2    &.03-0.05&   0.13-0.20  \\
M-0.15-0.07 & ...       & ...    &\le0.04 &   ...   \\
M+0.16-0.10 & 0.97-1.2  & 2.4    &.08-0.16&   0.23-0.48  \\
M+0.21-0.12 & 0.70-0.77 & 1.5    &.08-0.11&   0.15-0.21  \\
M+0.24+0.02 & \le 0.41  & 0.9    &\le 0.05&   0.08-0.09  \\
M+0.35-0.06 & \le 0.68  & 0.71   &.04-0.05&   0.13-0.19  \\
M+0.48+0.03 & \le 0.47  & \le0.42&...     &   \le 0.67  \\
M+0.58-0.13 & \le 0.43  & 0.73   &\le 0.10&   0.19-0.37\\
M+0.76-0.05 &...        & ...    &\le 0.45&   \le 0.69 \\
M+0.83-0.10 &...        & ...    &\le 1.0 &   \le 0.56 \\
M+0.94-0.36 &...        & ...    &...     &   ... \\
M+1.56-0.30 &...        & ...    &...     &   ...  \\
M+2.99-0.06 & \le 0.96  & \le0.71&\le 0.4 &  \le 0.21  \\
M+3.06+0.34 &...        & ...    &...     &   \le 0.63 \\
\noalign{\smallskip}
\hline
\end{tabular}
\end{table}

\begin{figure}
\includegraphics[bb=69 34 549 714, angle=270,width=8cm]{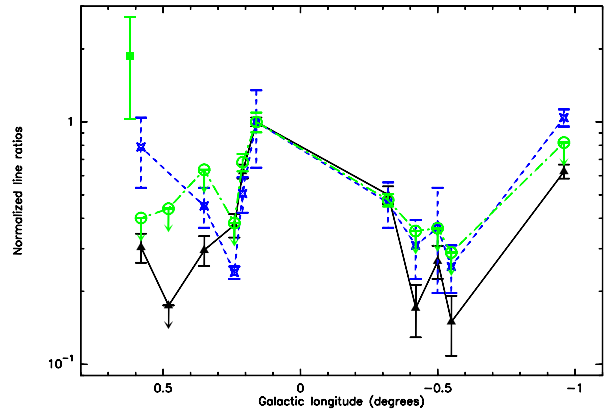}
\caption{ Line ratios as a function of the Galactic longitude (Triangles an solid lines: [\OIII]/[\NII]; crosses and dashed lines: [\OIII]/[SIII]; circles and dot-dashed lines: [\NIII]/[\NII]). All the line ratios have been normalized to the ratios measured for M+0.16-0.10. The solid  square and bars represent the [\NIII]/[\NII] ratio in the Sgr B envelope (Goicoechea et al. 2004)}
\label{fig_long}
\end{figure}

In addition, we have represented in Figure \ref{fig_long}  the [\OIII]/[\NII], [\NIII]/[\NII] and [\OIII]/[SIII]  line ratios  normalized to the values measured for M+0.16-0.10 as a function of the Galactic longitude.
 The expected errors in the line ratios due to the line fits errors and, mainly, to the extinction uncertainties are represented by the errorbars.
The sources with the highest excitation are M+0.16-0.10 and M+0.21-0.12, which are located in the Radio Arc region (Table \ref{tab_rat4}, Fig. \ref{fig_msx}). As discussed by Rodriguez-Fernandez et al. (2001a) the excitation of these sources is due to the combined effect of the Quintuplet and the Arches clusters, which are also responsible for the heating of the warm dust associated to The Sickle and the Thermal Filaments.
 The line ratios measured in M-0.32-0.19, M+0.24+0.02, M+0.35-0.06 and M+0.58-0.13 are $\sim 40$~\% of those measured in the the Radio Arc sources.
  Sources in the Sgr C region (M-0.42+0.13, M-0.55-0.05 and  M-0.50-0.03) present even lower line ratios ( 20\% of those of the Radio Arc ). Surprisingly, M-0.96+0.13 shows high line ratios (similar to the sources in the Radio Arc). For the rest of the sources no line  ratios sensitive to the effective temperature of the radiation can be derived, however the results published by Goicoechea et al. (2004) for SrgB2 imply that [\NIII]/[\NII] ~ ratios in the Sgr B2 envelope  (also shown in Figure \ref{fig_long}) are similar to those measured in the Radio Arc.
 Figure \ref{fig_long} shows local maximums  in the line ratios for the sources located at longitudes $l\sim$ 0.2, 0.6, -0.5 and -0.96 deg.
Hence, the excitation does not decrease smoothly with increasing distance to the known clusters of massive stars in Sgr A and the Radio Arc. Other ionizing sources should be present in the GC to account for the excitation of the ionized gas (see also Sect \ref{dis_medium}).

\begin{figure}[h!]
\includegraphics[angle=90,width=7.5cm]{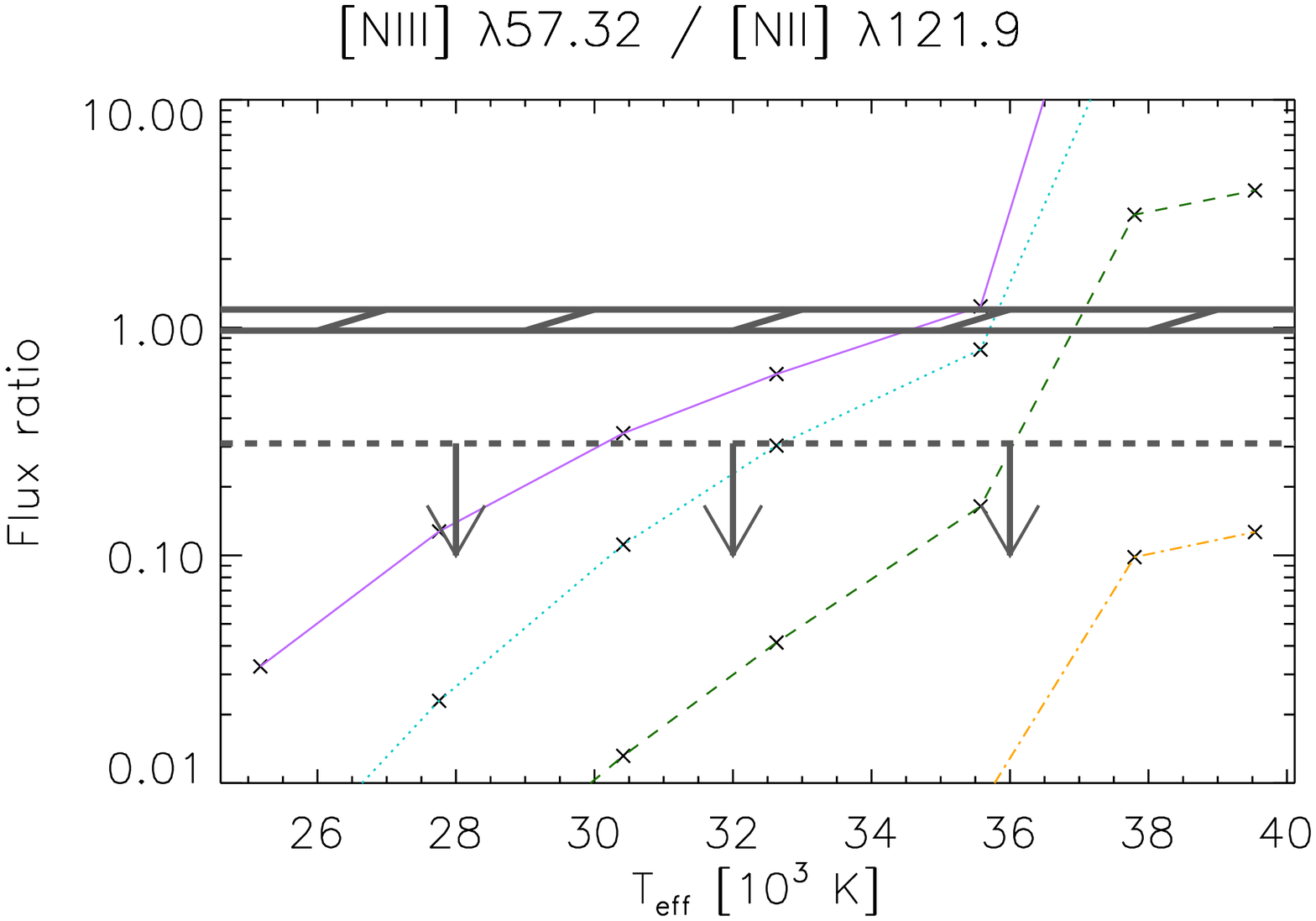}
\vspace{-5mm}

\includegraphics[angle=90,width=7.5cm]{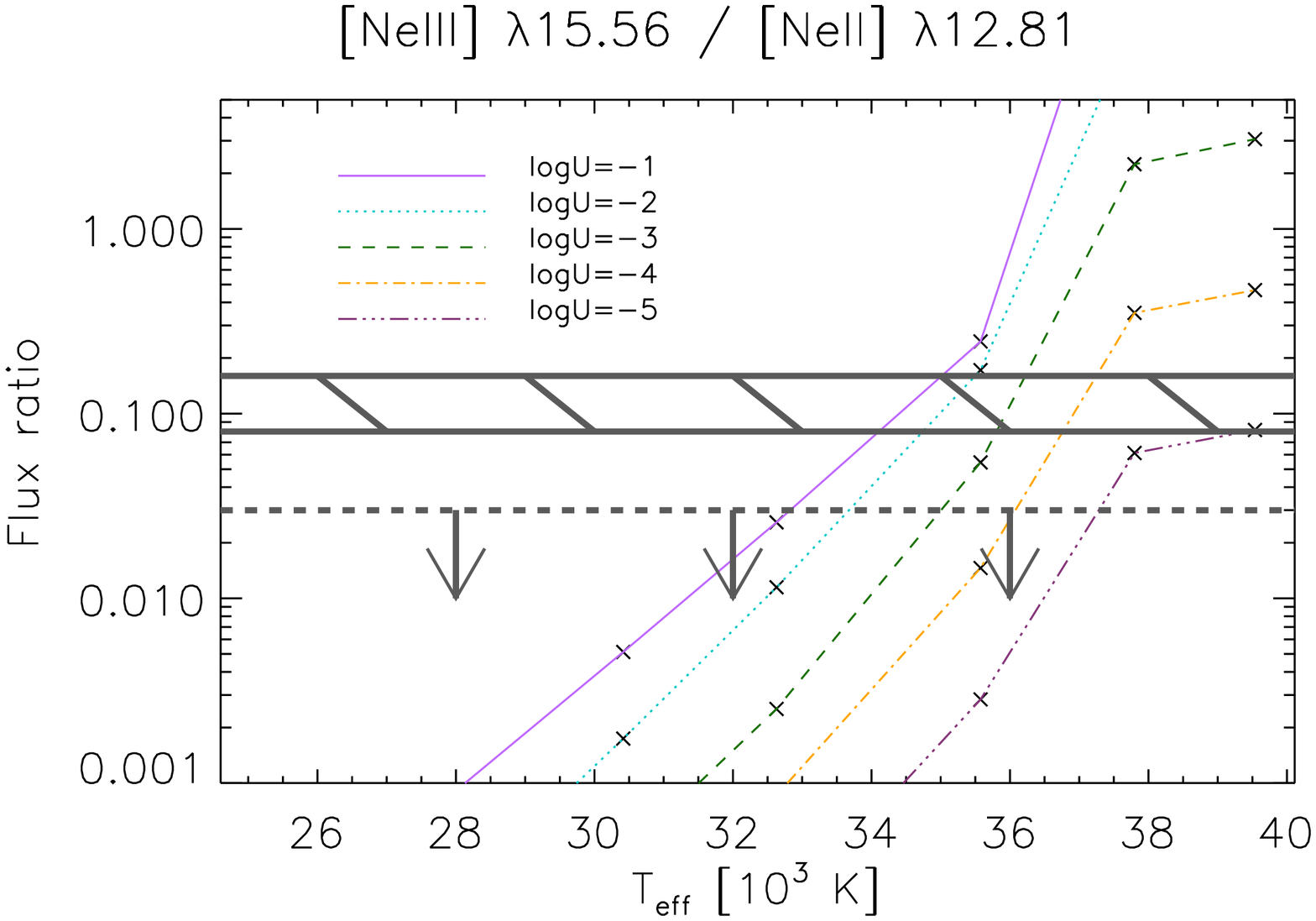}
\vspace{-5mm}

\includegraphics[angle=90,width=7.5cm]{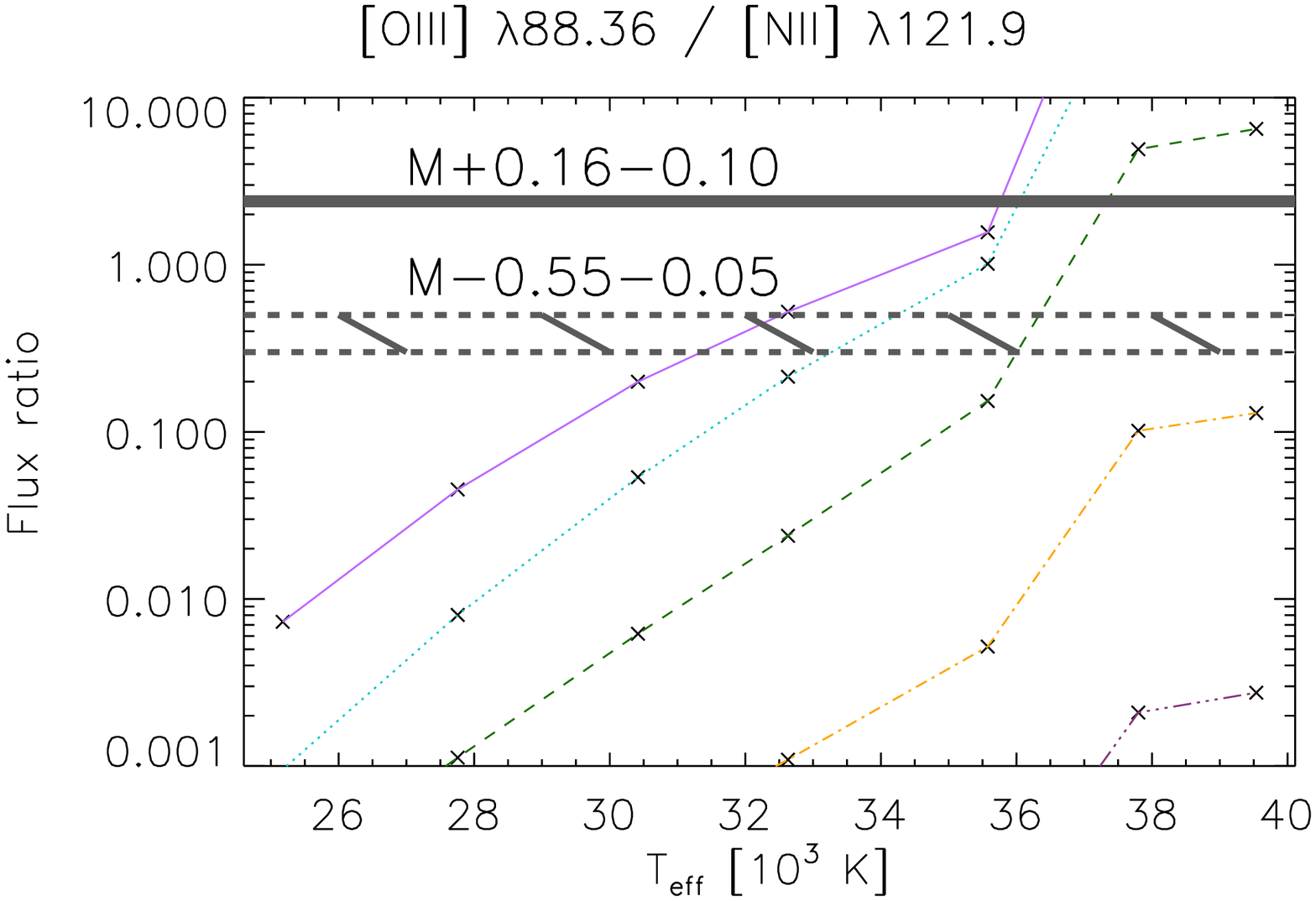}
\vspace{-5mm}

\includegraphics[angle=90,width=7.5cm]{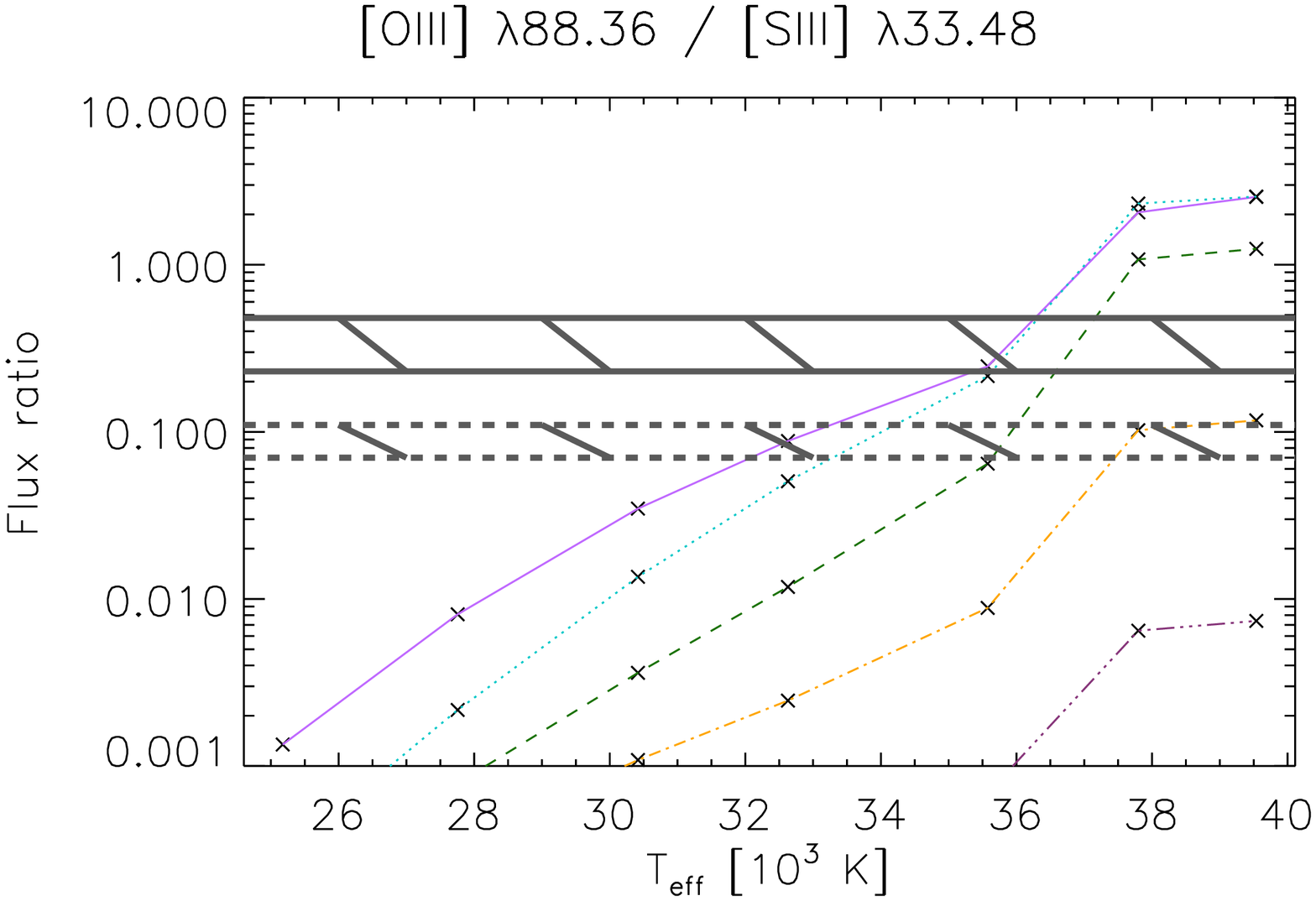}
\vspace{-3mm}

\caption{Fine structure line ratios modeled with MICE as a function of the ionization parameter ($U$) and the effective temperature of the ionizing radiation ($T_{eff}$). The measured ratios are shown with horizontal solid grey lines for the source with a highest excitation (M+0.16-0.10) and with grey dashed lines for the source with the lowest measured excitation (M-0.55-0.05). When only an upper limit can be derived the dashed line represents  that upper limit  (also indicated by arrows) }
\label{fig_cloudy}
\end{figure}

\subsection{CLOUDY modeling}
Apart from the sources of the Radio Arc, the morphology of the medium and the location of the hot stars are not known. Therefore, we cannot make a detailed study of the ionization for all the sources as we did in the Radio Arc region   (Rodríguez-Fernández et al. 2001b). Nevertheless, to get further insight into  the properties of the ionized gas, we have performed some simple photo-ionization model calculations  using the MPE IDL CLOUDY Environment (MICE), which was developed by H. Spoon at the Max-Planck-Institut f\"ur extraterrestische Physik (MPE) and uses CLOUDY 94 (Ferland 1996). To define the shape of the continuum radiation illuminating the nebulae we have taken the stellar atmospheres modeled by Schaerer $\&$ de Koter (1997). As nebular abundances we took  X(Ar)=-5.44 dex, X(O)=-3.00 dex, X(Ne)=-4.0 dex, X(S)=-4.60 dex and X(N)=-3.52 dex, in agreement with previous determinations in \HII ~ regions in the inner Galaxy (Simpson et al. 1995) and our new estimations. We have computed a grid of plane-parallel models in terms of the electron density ($n_e = 1-10^4$ \cmmt), the effective temperature of the radiation ($T_{eff}=25000-45000$ K) and the ionization parameter ($U=10^{-1}-10^{-5}$). The ionization parameter gives the number of photons with energy higher than 13.6 eV per hydrogen atom in the surface of the nebula. It can be expressed as

\begin{equation}
U=\frac{Q(H)}{4\pi D^2 n_e c}=\frac{\Phi(H)}{n_e c}
\label{eq_U}
\end{equation}

where $Q(H)$ is the Lyman continuum photons emission rate (\smu) emitted by the ionizing  source located at a distance D of the nebula and $\Phi(H)$ is the Lyman continuum photons flux (\smu \cmmd) in the nebula surface.

Figure  \ref{fig_cloudy} shows the predicted ratios as a function of $T_{eff}$ and $U$ for a nebular density of 10$^{1.5}$ \cmmt, which is the average density derived from our data (the results are very similar for any density in the 1-10$^3$ \cmmt~  range). For each line ratio, we have represented the measured range for the sources with the highest (M+0.16-0.10; solid lines) and lowest (M-0.55-0.05; dashed lines) line ratios in our sample. Note that in spite of the different beam sizes and uncertainties in the assumed nebular abundances, the range of $U$ and $T_{eff}$ needed to explain the  [\OIII]/[\NII] ~ and [\OIII]/[\SIII] ratios  in M+0.16-0.10 are in excellent agreement with those needed to explain the [\NIII]/[\NII] ~ and [\NeIII]/[\NeII] ~ ratios. It is also noteworthy that the line ratios measured for those extreme sources are close in the $U$ and $T_{eff}$ parameter  space. For a given $U$, $T_{eff}$ just changes by $\sim 2000$\,K from M-0.55-0.05 to M+0.16-0.10. All the measured ratios can be explained with effective temperatures, from 32000-35000 K if the ionization parameter is high ($\log U = -1 $) to 36000-37000 K if  $\log U = -3$, or even with higher $T_{eff}$ for lower ionization parameters.

Given the uncertainties on the possible ionizing sources, it is not easy to break the degeneracy in the $T_{eff}-U$ space. Fortunately for most of the sources where we can estimate the excitation ratios we have also detected the H\,{\sc i}~ Br$\alpha$ line. Therefore, we can estimate the flux of Lyman continuum photons ($\Phi(H)$) in the SWS beam from the Lyman continuum photons emission rate ($Q(H)$) derived from Br$\alpha$  (Eq. \ref{ccc}). With $Q(H)=10^{47}-10^{48}$ \smu ~ one can estimate a flux of Lyman continuum photons within the SWS beam of $10^{10-11}$ \smu \cmmd. Hence, using  Eq. \ref{eq_U}  and $n_e=10-100$\,\cmmt ~ one obtains log$U$ from -1 to -3. With these ionization parameters  one gets effective temperatures of the ionizing radiation of 35000-37000 K  for the high excitation sources and 32000-36000 K for the low excitation sources (Fig. \ref{fig_cloudy}).  The effective radiation temperature derived for the high excitation sources corresponds to stars with a spectral type between O9 and O7.5 and that of the low excitation sources is typical of stars with spectral type between B0.5 and O8   (Schaerer \& de Koter 1997).  However, there are three sources for which we could estimate the excitation but Br$\alpha$ has not been detected. These are M-0.96+0.13, M-0.55-0.05 and M+0.24+0.02. For these sources the non-detection of recombination lines imply upper limits to log$U$ between $-1.5$ and $-2.5$. In particular, given the high [\OIII]/[\SIII]~ and [\OIII]/[\NII] ~ ratios measured for M-0.96+0.13, one derives a lower limit to $T_{eff}$ of $\sim 35500 $ K.

\section{Discussion}

\subsection{Ionizing sources and the structure of the medium}
\label{dis_medium}

The  main ionizing source of M+0.21-0.12 and M+0.16-0.10, which are located in the Radio Arc region, is certainly the Quintuplet cluster.  Indeed, Rodríguez-Fernández et al. (2001a) have shown that the combined effects of the Quintuplet and the Arches clusters ionize a large region of more than 30$\times$30 pc$^2$.
The density of the observed nebulae in this  region is $\geq 100$ \cmmt. However, CLOUDY model calculations showed that  the medium cannot have an homogeneous density of $\sim 100$ \cmmt ~ in order to explain the large size of the ionized region (which otherwise will not be larger than $\sim 6$ pc). Some sort of non-homogeneities (gas with a density lower than 10 \cmmt) should exist in between the stars and the dense nebulae (Rodríguez-Fernández et al. 2001a).

As mentioned in Sect. \ref{sect_cloudy}, the known clusters of massive stars cannot account for the excitation of the fine structure lines over  the whole GC region. There must be more ionizing sources. The observed clouds were selected far from the thermal continuum sources (typically at 10-45 pc).  However, it is  possible that they are ionized by the same sources that ionize the \HII ~ regions and heat the dust in the prominent continuum complexes Sgr C, B, D,... For instance, M+0.58-0.13 could be ionized  by the sources that ionize the Sgr B region. M-0.50-0.03 and M-0.55-0.05 may be ionized by the sources in Sgr C. M-0.32-0.19 could be ionized by the source that heat the dust at $(l,b)\sim(-0.35^\circ,-0.2^\circ$). In contrast, it is not clear where the possible ionizing source of  M-0.96+0.13 could  be located since both the Sgr C and the Sgr E complexes are rather far ($> 40$ pc) from this source.

We have undertaken CLOUDY simulations to estimate the size of the ionized region around an ionizing source with a $T_{eff}$ of 35500 K assuming an homogeneous medium with a density of 10 \cmmt ~ (typical of the sources other than those located in the Radio Arc, see Sect. 5). We have found that the ionized region radius is $\sim 11$ pc for a Lyman continuum photons emission rate ($Q(H)$) of 10$^{49}$ \smu, while it is increased to 25 and 53 pc for $Q(H)$ values of 10$^{50}$ \smu  and 10$^{51}$ \smu, respectively. These results  are weakly dependent (less than 10 \%) on the value of $T_{eff}$ in the range 27000 -44000 K. On the other hand, those sizes would increase by a factor of $\sim 5$ if the gas density is 10 times lower.

It is difficult to estimate the actual rate of Lyman continuum photons escaped from the radio continuum sources. Adding the values derived by different authors (Gaume et al. 1995; Liszt 1992; Liszt \&  Spiker 1995)  for the discrete \HII~ regions one finds a total $Q(H)$ of $\sim 10^{50.3}$ \smu~ in Sgr B2, $\sim 10^{49.8}$ and $\sim 10^{49.5}$ \smu~ in Sgr D and C, respectively, and a somewhat lower value ($\sim 10^{49}$ \smu) for the Sgr E complex. On the one hand, in the Galaxy only around 16 \% of the O stars produce discrete \HII~ regions (Mezger 1978). On the other hand, as mentioned in Sect. 1, Mezger \& Pauls (1979) have estimated a contribution to the thermal radio continuum in the GC of 40 \% from discrete \HII ~ regions  and 60 \% from the extended low density ionized gas. Therefore, one would expect that the number of photons that ionize the diffuse gas is within 1 and 5 times that measured in the discrete \HII~ regions.

A $Q(H)$ of $\gsim 10^{50}$ \smu ~ arising from Sgr B, C or D could thus explain the ionization of clouds located at 10-20 pc if the density is not higher than 10 \cmmt. On the contrary, up to $Q(H)=10^{51}$ \smu ~ escaping from the Sgr E region would be needed to explain the presence of ionized gas in M+0.96+0.13, which is located at more than 40 pc from this complex. However, this $Q(H)$ is at least a factor of 10 higher than that expected for Sgr E. To explain the ionization of M+0.96+0.13 as due to Sgr E with a more plausible $Q(H)$ of 10$^{49}-10^{50}$ \smu ~ the medium cannot have an homogeneous density of 10 \cmmt ~ but should present non-homogeneities with lower  density by at least a factor of 10.

An independent piece of evidence for   the structure of the GC interstellar medium and the origin of the ionizing radiation is the apparent discrepancy between the effective temperatures and the Lyman continuum photons emission rate derived in the previous sections. The Lyman continuum photons rate emitted by a star  with the effective temperature derived in Sect. \ref{sect_cloudy} is at least 10 times higher than that measured in Sect. \ref{sect_lyc} (see Schaerer \& de Koter 1997). Indeed the Lyman continuum photons rate emitted by the ionizing sources could be much higher if the ionizing radiation comes from a star cluster instead of a single star. The apparent inconsistency between effective temperatures and the number of ionizing photons implies that the ionizing sources are located far from the nebulae. The distance from the ionizing sources to the nebulae can be estimated assuming a simple geometry. For instance, if the  ionized gas emission comes from a shell of radius $r$ and the ionizing source (located at the center of the shell) emits $Q_T$ Lyman continuum photons per second, then the Lyman continuum photons flux in the shell surface would be $Q_T/(4 \pi d^2)$. Assuming that the shell is located at the distance of the GC, $D$, the number of ionizing photons within a beam $\Omega$, $Q_\Omega$, will be approximately given by:

\begin{equation}
 Q_\Omega = \frac{Q_T}{4 \pi r^2} \Omega D^2
\end{equation}

and thus, taking a beam of 20$^{''}$ and a distance $D=8.5$~kpc one obtains:

\begin{equation}
 r(\mathrm{pc})=0.24 \sqrt{Q_T/Q_\Omega}
\end{equation}

Let us consider that one only measures 10$^{47.3}$ photons  per second of those emitted by an O8 star  (10$^{49}$ \smu, Schaerer \& de Koter 1997). Then $Q_T/Q_\Omega=50$ and the ionizing source would be located at $\sim$ 1.5 pc from the nebula. If the ionizing source is a cluster instead of a single star, $Q_T$ and $r$ would be larger. Therefore, the typical spatial scale of the 
non-homogeneities is probably larger than 1.5 pc.  Indeed, large 
non-homogeneities  seem to be  a general characteristic of the dense gas in the GC. About  300 molecular shells with sizes of between 5 and 10 pc have been identified from the CO survey of the GC by Hasegawa et al. (1998). Furthermore, hot (60-130 K) expanding shells of molecular gas with sizes of 1-2.5 pc (consistent with the sizes required to explain the ionized gas) have been observed in the envelope of the Sgr B2 molecular cloud. Wind blown bubbles driven by evolved massive stars are the most likely explanation (Martin-Pintado et al. 1999).

In conclusion, the properties of the ionizing radiation in the 400 central pc of the Galaxy seem to be rather constant  (a medium permeated by radiation from relatively hot and distant stars) and similar to that arising from the massive star clusters in the Radio Arc.

\subsection{Comparison with external Galaxies}

\subsubsection{Ionization state and starburst properties}
 As shown in the previous sections, the nebulae are ionized by relatively hot but distant stars and the medium, at least in some regions of the GC, is clearly non-homogeneous. As a consequence, the radiation is diluted and the ionization parameter is lower than log$U=-1$ (and higher than log$U=-3$). These properties are similar to those found in starburst  galaxies. In the detailed analysis of M82 by Scheiber et al. (1999), they have derived an homogeneous ionization parameter, log$U=-2.3$, over the 450 central pc of the galaxy. They have estimated effective distances from the ionizing sources to the ionized nebulae of $\sim 25$ pc. The ionization parameter derived for NGC253 and NGC3256 by Thornley el al. (2000) are log$U=-2.6$ and log$U=-2.3$.
 As discussed by Carral et al. (1994) and Thornley et al. (2000) non-homogeneity is a typical characteristic of the ISM in starburst galaxies and it is independent of the global luminosity and the different possible triggering mechanisms of the starburst. Therefore, the non-homogeneity of the medium should be inherent to the star formation activity itself. For instance, it can be due to the  interaction of massive stars, in particular in the last stages of their life, with the surrounding medium. As discussed in the previous section, this is the most likely explanation in the case of the GC.

 Regarding the effective temperature of the ionizing radiation, it  is $\sim 32000-37000$  K in the GC (Sect. \ref{sect_cloudy}) . These values are lower than those measured in luminous infrared galaxies like NGC 3256 or NGC 4945 ($\sim 40000$ K, Lutz et al. 1996). In contrast, they are in good agreement with those derived in lower luminosity starburst as Cen A, M82 or M83 (34500-35500 K, Unger et al. 2000).  Indeed, moderate excitation ratios and effective temperatures are common in starburst galaxies. In the sample of 27 starburst galaxies studied by Thornley et al. (2000) the [\NeIII]15.5/[\NeII]12.8 ratio varies from 1-3 for NGC55 or NGC5253 to  0.05-0.1 for galaxies as IC342, NGC 253, NGC1482 or M83. The former value is in perfect agreement with that measured in the central pc of the Milky Way ( 0.07; Lutz et al. 1999) and the GC sources presented in this paper ( Table \ref{tab_rat4}). The detailed analysis by Thornley et al. (2000) of the gas photo-ionization taking into account the evolution of the stars in the HR diagram, showed that the low excitation measured in those galaxies does  not imply intrinsic differences in the star formation activity as a different upper mass cutoff of the initial mass function (IMF). On the contrary, the low ratios are most probably due to  starbursts  of short duration (a few Myr) that produce hot massive stars but whose age quickly softens the radiation and the diagnostic ratios.
 For instance, their Fig. 10 shows how the [\NeIII]/[\NeII] ratio decreases with time after a short burst of 1 Myr with an upper mass cutoff of 100 M$_\odot$.
For an ionization parameter of $\log U=-1.5$ the  [\NeIII]/[\NeII]  ratio decreases from $\sim 20$ to 0.05 in $\sim 8$ Myr, while for an ionization parameter of $\log U=-2.3$ that line ratio decreases from $\sim 4$ to 0.05 in $\sim 6$ Myr.
Therefore, the line ratios measured in the GC are consistent with a short burst of massive star formation that took place less than 8 Myr ago.
This is in agreement with the properties of the stars studied in the three known clusters of massive stars located in the GC (Figer 2004).

\subsubsection{NeII versus far-infrared luminosity} 

\begin{figure}[tbh]
\includegraphics[bb=166 47 523 416,angle=-90, width=7.5cm]{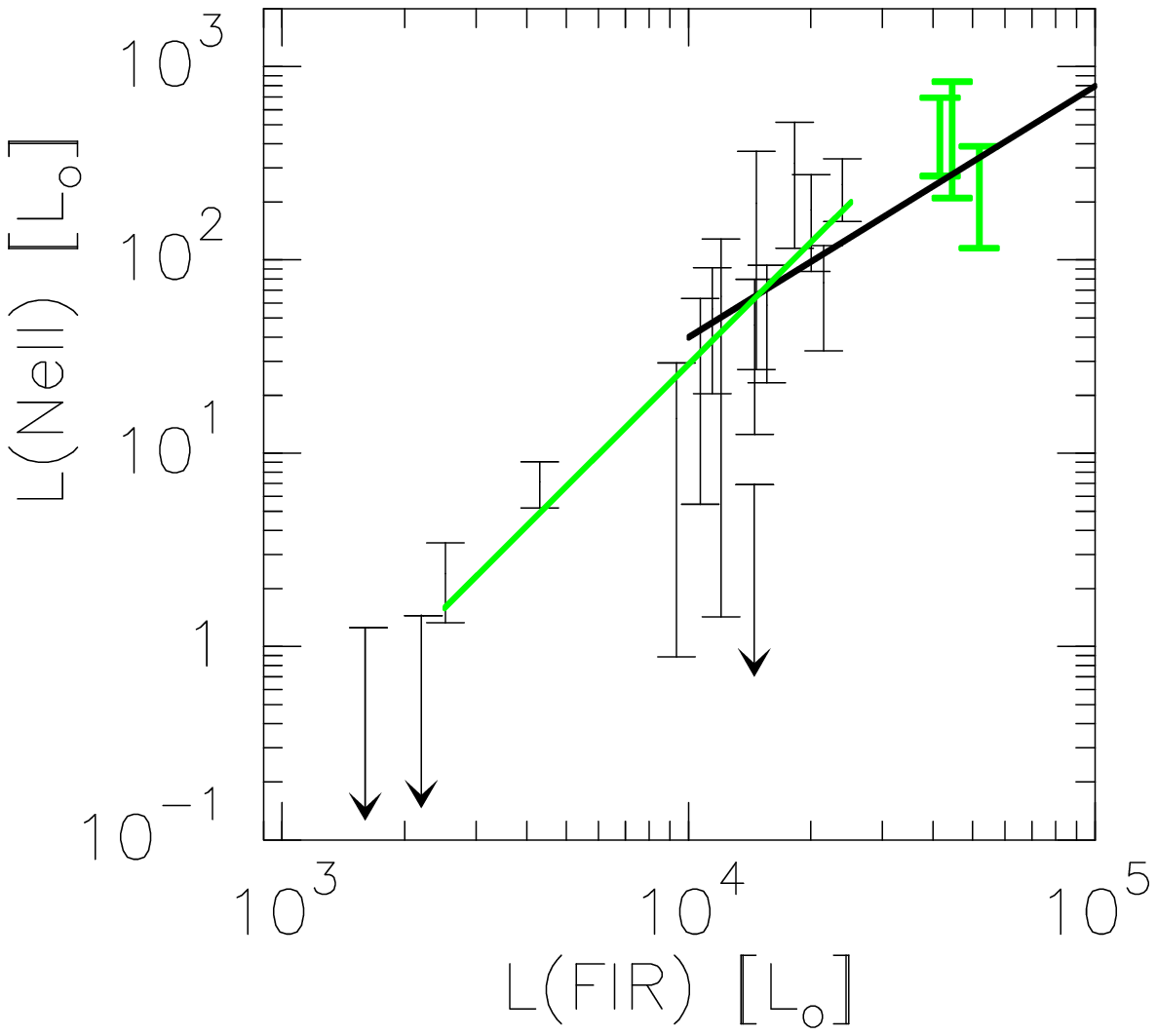}
\caption{[\NeII] 13 \mum~ luminosity versus the FIR luminosity for the GC molecular clouds. Thin black bars represent the results for the source sample of Table. 1. The arrows represent upper limits to the [\NeII] luminosity. Thick grey bars represent the results for G0.18-0.04, G0.02-0.07 and Sgr C. The length of the bars account for the extinction uncertainties. Black and grey lines represent different fits to the data (see text).}
\label{fig_neii_fir_gc}
\end{figure}

Figure \ref{fig_neii_fir_gc} shows the [\NeII] 13 \mum ~  luminosity versus the far-infrared (FIR) luminosity  for the GC clouds. The FIR luminosity  has been obtained from the LWS continuum spectra (Rodríguez-Fernández et al. 2004) by applying a correction factor corresponding to the ratio of the SWS and the LWS apertures  (the dust luminosities uncertainties are $\sim 10$ \%).  The thin bars account for the extinction uncertainties  while the arrows represent the upper limits to the [\NeII] luminosity for the sources where the  line has not been detected.  The [\NeII]  luminosity exhibits a clear trend to increase with an increasing dust continuum luminosity. This suggests that the dust heating should be related to the ionizing source of the gas.  We have fitted a line to the data for the sources of our cloud sample with detected [\NeII] emission. As a representative value for the [\NeII] luminosity of those sources we have taken the medium value between those obtained in the upper and lower limits to the extinction (the range between the medium and the upper and lower values has been considered as the error of the measure and has been taken into account to do the fits). The fit result is represented by a grey straight line in Fig. \ref{fig_neii_fir_gc} and can be expressed as:

\begin{equation}
\log(L_{NeII})=(2.0\pm0.3)\log(L_{FIR})-(6.7\pm1.0)
\label{eq_fit}
\end{equation}

\begin{figure}[tbh]
\includegraphics[angle=-90,width=8.5cm]{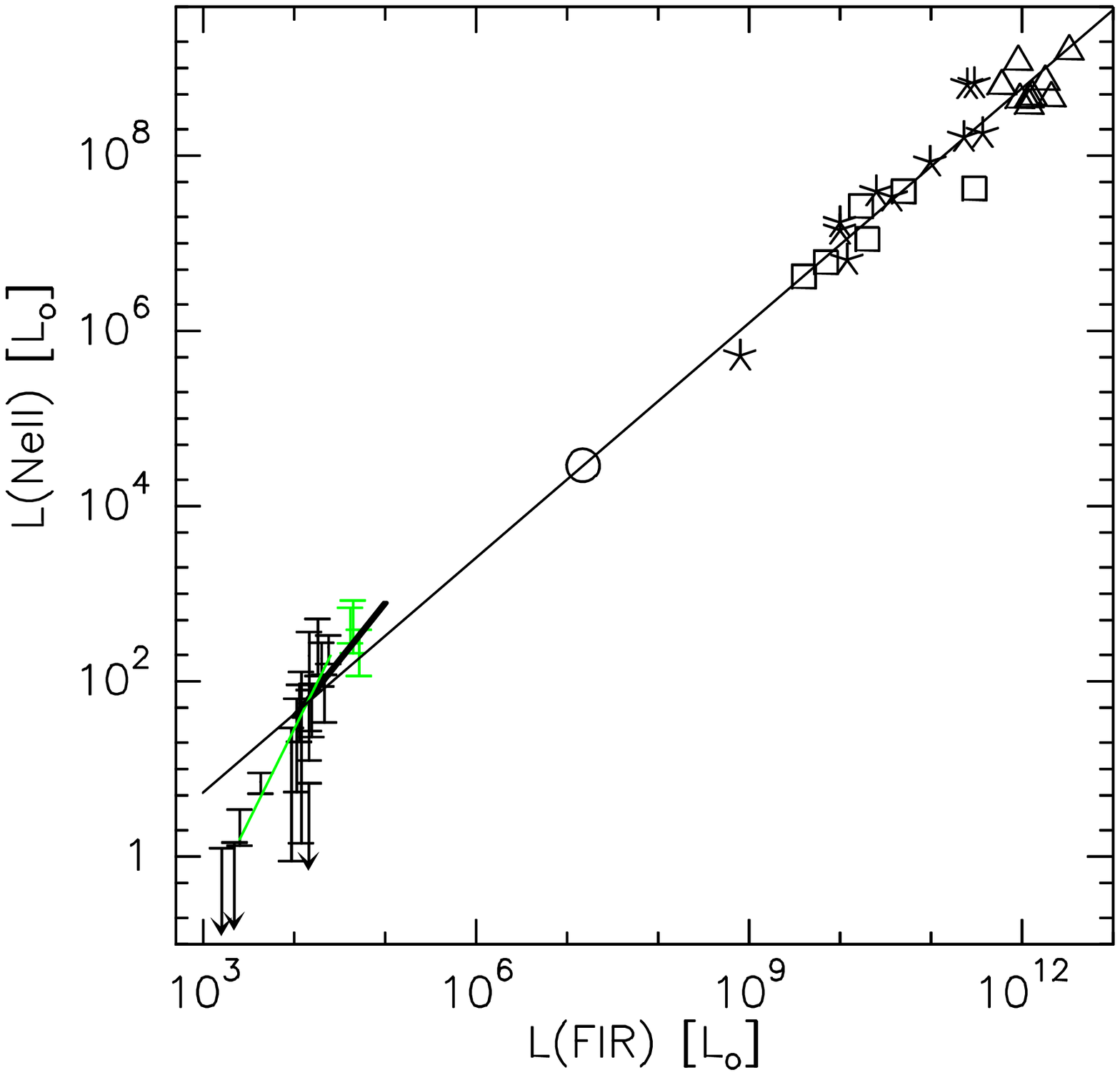}
\caption{[\NeII] 13 \mum~ luminosity versus the FIR luminosity for external galaxies and GC clouds. The GC clouds are represented by bars that account   for the extinction uncertainties. The other markers represent results from Genzel et al. (1998) for the central parsecs of the Galaxy (empty circle), galaxies with an active nucleus (AGNs; empty squares), Ultra Luminous Infrared Galaxies (ULIRGs; triangles) and starburst galaxies (stars). The thin black line is a fit to the Genzel et al. Data. The thick black line and the grey line are different fits to the GC data (see text and Fig. \ref{fig_neii_fir_gc})}
\label{fig_neii_fir}
\end{figure}

Genzel et al. (1998) and Sturm et al. (2002) have also found a [\NeII]  to FIR correlation in external galaxies. Figure \ref{fig_neii_fir} shows the [\NeII]  to FIR plot done with  data of starburst, AGNs and Ultraluminous Infrared Galaxies (ULIRGs) from Genzel et al. (1998).  We have fitted a line to these data obtaining:

\begin{equation}
\log(L_{NeII})=(0.90\pm0.13)\log(L_{FIR})-(1.5\pm0.5)
\label{eq_fit2}
\end{equation}

 The  line is plotted as a black thin line in Fig. \ref{fig_neii_fir}, which also shows the GC data of Fig. \ref{fig_neii_fir_gc}.
 Prolongating this line down to FIR luminosities of 10$^3$ L$_\odot$, one finds that the  [\NeII]-to-FIR ratios measured for the GC sources (except for the four clouds with lowest luminosities) are in agreement with those derived from the external galaxies data.
However, the slope of the line displayed in Eq. \ref{eq_fit2}  is half of that in Eq. \ref{eq_fit},
 which is also shown in Fig. \ref{fig_neii_fir} as a thin grey line.
The discrepancy is probably due to the fact that the outer GC sources (those located in the extremes of the Central Molecular Zone and in the Clump 2) present a lower [\NeII] 13 \mum ~ to continuum ratio than the inner sources.
The reason can be that for the outer sources a significant fraction of the FIR continuum arises in dust heated by stars that do not ionize the neon.
To get further insight in the [\NeII]-to-FIR correlation we have expanded the luminosity range of the GC sources. We have retrieved from the ISO Data Archive observations of three well-known sources as G0.02-0.07, the Sickle (G0.18-0.04) and Sgr C. The data have been reduced like the other GC clouds (Sect. 2) and corrected for extinction in the range $A_V=30-60$ mag. These three sources are represented in Figs.\ref{fig_neii_fir_gc} and  \ref{fig_neii_fir} by thick grey bars. We have fitted a line to the data for inner GC sources (those located at galactic longitudes between -0.55 deg and +0.35  deg, including the three sources retrieved from the ISO Archive). The fit is shown as a thick black line in  Figs. \ref{fig_neii_fir_gc} and \ref{fig_neii_fir} and can be expressed as:

\begin{equation}
\log(L_{NeII})=(1.3\pm0.3)\log(L_{FIR})-(3.5\pm1.5)
\label{eq_fit3}
\end{equation}

The parameters of this line are in good agreement (within errors) with those of the fit to the external galaxies data (Eq. \ref{eq_fit2}). On the opposite, they differ significantly with respect to those obtained including the outer GC sources (Eq. \ref{eq_fit}). Therefore, we conclude that the fraction of the FIR emission that is due to dust heated by non-ionizing stars is higher for the outer than for the inner GC sources.

Sturm et al. (2002) have proposed that the [\NeII]-to-FIR correlation is important to understand the origin of the far-infrared (FIR) emission in active galaxies. Both the AGN and star formation have been proposed. Taking into account that the [\NeII] line  is a good tracer of hot stellar radiation, they suggest that the correlation between the [\NeII] line and the FIR continuum in starburst and active galaxies implies a common origin due to stellar radiation. However, starburst and AGN activities are commonly associated and the SWS beam is rather large for detailed studies of external galaxies. Therefore, Sturm et al. do not  exclude scenarios in which the FIR luminosity of AGNs would be  powered by the active nuclei (the mid-infrared continuum emission is indeed dominated by the dust continuum emission of the black hole accretion disk). The GC sources have been selected as standard molecular clouds  and both the  [\NeII] 13 \mum ~  and the FIR continuum emissions are due, without doubt,  to stellar radiation. Therefore, the [\NeII] to FIR ratio measured in the inner GC clouds supports the scenario proposed by Sturm et al. (2002) and confirms that no contribution from the AGN is needed to explain the FIR emission in active galaxies.

\section{Summary and conclusions}
We have presented fine structure and recombination line  observations made with
ISO toward a sample of 18 sources located in the central 500  pc of the Galaxy.
The sources were selected as molecular clouds located far from thermal radio continuum and FIR sources. The main results can be summarized as follows:

\begin{itemize}
\item Fine structure lines from [N{\sc ii}] and [S{\sc iii}] (excitation potential higher than 20 eV) have been detected in 16 sources. In 10 sources we have even detected [O{\sc iii}] lines (excitation potential higher than 30 eV).

\item We have used several techniques to determine lower ($\sim 25$) and upper ($\sim 60$) limits to the extinction to correct the observed line fluxes.

\item  Nitrogen, neon, and sulfur abundances have been derived for 8 sources. The neon abundance is similar  to that of \HII~ regions in the 5-kpc ring and in starburst galaxies while the sulfur abundance is higher in the GC than in the 5-kpc ring probably due to a lower depletion.

\item The excitation of the fine structure lines has been studied in 10 sources. Line ratios imply effective temperatures of the ionizing radiation of 32000-37000 K and  ionization parameters, $U$, of  $-1>\log U>-3$.

\item The highest excitation is found in the Radio Arc region but it does not decrease smoothly with distance. The known clusters of massive stars in the GC are not enough to account for the ionization over the 500 central pc. Although the clouds are relatively far from the main radiocontinuum complexes (10-30 pc, and up to 45 for M-0.96+0.13), it is possible  that they are ionized by escaped photons  from those complexes. 

\item The comparison of the effective temperatures of the ionizing radiation to the measured Lyman continuum photons emission rate imply  that the clouds are ionized by distant sources.

\item The properties of the radiation and the interstellar medium in the GC are similar to those found in low excitation starburst galaxies. The relatively low excitation in the GC is consistent with a short burst of massive star formation $\sim 7$ Myr ago. The radiation from the hot massive stars has softened due to aging effects.

\item We have found a [\NeII] to FIR continuum ratio for the GC clouds similar to that found in external galaxies. In the GC clouds, both the  [\NeII] and FIR  emission arises unambiguously from stellar radiation. Therefore, this finding supports the idea of Sturm et al. (2002) that the FIR continuum in Active Galaxies is associated to dust heated by stellar radiation and not by the AGN.

\end{itemize}

\begin{acknowledgements}
We thank an anonymous referee for his/her critical comments that have helped us to improve the paper and Tom  Spain for his help with the English. NJR-F acknowledges useful comments from A. Verma and M. Morris. NJR-F has been supported by a Marie Curie Fellowship of the European Community program ``Improving Human Research Potential and the Socio-economic Knowledge base'' under contract number HPMF-CT-2002-01677. The authors have been partially supported by the Spanish {\it Ministerio de Ciencia y Tecnologia} (MCyT) grants AYA2002-10113-E, AYA2003-02785-E and ESP2002-01627. This  research has made use of data products from the Midcourse Space Experiment. Processing of the data was funded by the Ballistic Missile Defense Organization with additional support from NASA Office of Space Science. The data was  accessed by services provided by the NASA/IPAC Infrared  Science Archive. MICE, SWS and the ISO Spectrometer Data Center at MPE are supported by DLR (DARA) under grants 50 QI 86108 and 50 QI 94023.
\end{acknowledgements}

\end{document}